\documentclass{aa}  

\usepackage{graphicx}
\usepackage{txfonts}
\usepackage{graphicx}	% Including figure files
\usepackage{amsmath}	% Advanced maths commands
\usepackage{subfigure}
\usepackage{subcaption}
\usepackage{hyperref}
\usepackage{natbib}
\usepackage[dvipsnames]{xcolor}
\usepackage[normalem]{ulem}
\setcitestyle{notesep={; }}

\begin{document}

   \title{Approaching ballistic motion in 3D simulations of gamma-ray burst jets in realistic binary neutron star merger environments}

   \author{E. Dreas
          \inst{1,2,3}\thanks{E-mail: edreas@sissa.it}
          \and
          A. Pavan\inst{2,4}\fnmsep
          \and
          R. Ciolfi\inst{2,4}\thanks{E-mail: riccardo.ciolfi@inaf.it}
          \and
          A. Celotti\inst{1,3,5,6}
          }

   \institute{SISSA, Via Bonomea 265, I-34136 Trieste, Italy            
        \and
             INAF, Osservatorio Astronomico di Padova, Vicolo dell'Osservatorio 5, I-35122 Padova, Italy
        \and
        INFN, Sezione di Trieste, Via Valerio 2, I-34127 Trieste, Italy
        \and
        INFN, Sezione di Padova, Via Francesco Marzolo 8, I-35131 Padova, Italy
        \and
        INAF, Osservatorio Astronomico di Brera, via Bianchi 46, I-23807 Merate, Italy
        \and
        IFPU, Via Beirut 2, I-34151 Trieste, Italy
             }

   \date{Received 29 July 2024; accepted 11 January 2025}

% \abstract{}{}{}{}{} 
% 5 {} token are mandatory
 
\abstract 
{
The concomitant observation of gravitational wave and electromagnetic signals from a binary neutron star (BNS) merger in 2017 confirmed that these events can produce relativistic jets responsible for short gamma-ray bursts (sGRBs). The complex interaction between the jet and the surrounding post-merger environment shapes the angular structure of the outflow, which is then imprinted in the prompt and afterglow sGRB emission.}
{The outcome of relativistic (magneto)hydrodynamic simulations of jets piercing through post-merger environments is often used as input to compute afterglow signals that can be compared with observations. However, for reliable comparisons, the jet propagation should be followed until nearly ballistic regimes, in which the jet acceleration is essentially over and the angular structure is no longer evolving. This condition is typically reached in 2D simulations, but not in 3D ones. Our goal is to extend a (specific) jet simulation in 3D up to a nearly ballistic phase and analyse the overall dynamical evolution from the jet breakout.}
{Our work is based on a previous 3D magnetohydrodynamic jet simulation employing a realistic environment imported from a BNS merger simulation, extended here far beyond the evolution time originally covered. After approximately 3 seconds of the jet evolution on the original spherical grid, we remapped the system into a uniform Cartesian grid and reached about 10 seconds without loss of resolution.}
{The specific jet considered here struggled to pierce the dense surroundings, resulting in a rather asymmetrical emerging outflow with a relatively low Lorentz factor. 
Analysis of the energy conversion processes and corresponding acceleration showed that at the end of our simulation, 98\% of the energy is in kinetic form. Moreover, at that time the angular structure is frozen. We thus obtained suitable inputs for computing the afterglow emission. Our procedure is general and applicable to any jet simulation of the same kind.}
   {}

   
   \titlerunning{Ballistic GRB jets from BNS mergers}
\authorrunning{E.~Dreas et al.}
   \maketitle
%%%%%%%%%%%%%%%%%%%%%%%%%%%%%%%%%%%%%%%%%%%%%%%%%%

%%%%%%%%%%%%%%%%% BODY OF PAPER %%%%%%%%%%%%%%%%%%

\section{Introduction}
\label{sec:intro}

Gamma-ray bursts (GRBs) are bright explosive events characterised by an early prompt emission phase detected in the sub-MeV band followed by a shallower radio-to-TeV afterglow emission phase that can last for up to days, months, or even years after the initial burst. 
Numerous investigations have indicated that both the prompt and afterglow emission arise from the energy dissipation of a collimated relativistic outflow (`jet') launched from a compact central engine (e.g. a hyper-accreting black hole or a highly magnetised fast-spinning neutron star). The jet propagates through the surrounding medium, generating shocks and triggering various dynamical effects (including turbulence and magnetic reconnection) and leading to radiative dissipation through the synchrotron and inverse-Compton processes \citep[e.g.][]{Kumar2015}. 
These investigations have further suggested that the progenitor system responsible for the formation of the central engine powering the jet is either a core-collapsing massive star or a merging compact binary system \citep[e.g.][]{ZhangBing2018}. 
In particular, the association between at least a fraction of `short' GRBs (sGRBs; defined as those with a prompt emission phase lasting less than 2\,s) and binary neutron star (BNS) mergers has been recently confirmed by the concomitant observation of gravitational waves (GWs) from a coalescing BNS (event named GW170817) and a short duration burst of gamma rays (GRB\,170817A), with the latter detected $\mathrm{\simeq\!1.74}$\,s after the estimated time of merger \citep{LVC-BNS,LVC-MMA,LVC-GRB}. 
Analyses of the afterglow emission accompanying GRB\,170817A have further revealed that a collimated relativistic jet, launched from the merger remnant and propagating along a direction misaligned with respect to Earth, is responsible for the electromagnetic (EM) emission, and it has imprinted its angular structure and energetics onto the observed signals \citep[][]{Mooley2018b,Mooley2018c,Lazzati2018,Ghirlanda2019}. 
Providing conclusive evidence that BNS mergers can launch collimated relativistic outflows and power GRBs, such a discovery has promoted unprecedented efforts in exploring this connection \citep[e.g.][]{Ciolfi2020c,Margutti2021}.

The angular structure of a sGRB jet from BNS coalescence is shaped during the launch and the subsequent propagation through the post-merger environment surrounding the compact central engine \citep[][and refs.~therein]{Salafia2022}. 
After breaking out of such an environment, the jet angular structure is expected to still evolve for a while until it eventually freezes out. 
At the same time, the jet keeps accelerating until essentially all the available energy is in kinetic form. When those conditions are reached (ballistic regime), the jet expands freely in the interstellar medium (ISM) on orders of magnitude longer time scales until the interaction with the latter starts having relevant dynamical effects and the afterglow radiation is produced.

The afterglow signal is therefore expected to carry key information about the jet launching and the following propagation that is complementary to what can be inferred from GW signals.
In order to exploit such information, it is necessary to accurately model the jet propagation dynamics, and in this context, numerical simulations plausibly represent the most promising approach. 

Since the GRB\,170817A event, the evolution of sGRB jets from BNS mergers has been extensively investigated through relativistic hydrodynamic or magnetohydrodynamic simulations, performed in two or three spatial dimensions \citep[e.g.][]{Lazzati2017,Murguia2021,Geng2019,Kathirgamaraju2019,Nathanail2020,Nathanail2021,Urrutia2021,Urrutia2023,Gottlieb2020,Gottlieb2021,Gottlieb2022a,Gottlieb2022b,Gottlieb2023a,Garcia2023}.
While most of these simulations employ hand-made surrounding environments to study the propagation of incipient jets, the first investigations based on more realistic environments directly imported from the outcome of BNS merger simulations were recently accomplished, both in the absence and in presence of magnetic fields \citep{Pavan2021,Pavan2023,Lazzati2021}.\footnote{see also \cite{Nativi2021,Nativi2022}}

The outcome of these simulations, in terms of jet energetics and angular structure, can in principle be used as input to compute the corresponding multi-wavelength afterglow signals, depending on the viewing angle; the density of the ISM; and other parameters (e.g. \citealt{Lazzati2018}).
However, the obtained light curves are reliable only if the jet expansion has reached a nearly ballistic regime. Such a condition is typically met in studies based on two-dimensional simulations (e.g. \citealt{Lazzati2018,Xie2018,Urrutia2021}) but not in the case of three-dimensional simulations (e.g. \citealt{Kathirgamaraju2019,Nathanail2021,Nativi2022}).

In this work, we present the methods and procedure that we developed to extend, without loss of resolution, simulations such as the one presented in \citet[][P23 henceforth]{Pavan2023} up to a nearly ballistic jet expansion phase.
Taking the simulation of P23 as a reference model, we show how the evolution can be extended up to $\approx\!10$\,s after jet launching, when its angular structure is frozen and its energy is $\approx\!98$\% kinetic.
Besides demonstrating the methods, we study the dynamics and energetics of the specific jet examined, and we extract the final angular profiles of energy and Lorentz factor that can be employed as input to compute the corresponding afterglow light curves.

The paper is structured as follows. In Section \ref{sec:model}, we describe the numerical setup of the reference jet simulation, along with the methods implemented to achieve the extended evolution up to a nearly ballistic phase and to define the jet `head' for the following analysis. 
In Section \ref{sec:results}, we present our results in terms of jet dynamical evolution and energetics. In Section \ref{sec:angular}, we perform a detailed study of the jet structure and the evolving angular distribution of energy and Lorentz factor. 
In Section \ref{towards-afterglow}, we estimate the saturation Lorentz factor and deceleration radius for the jet model at hand.
Finally, in Section \ref{sec:conclusion}, we summarise the present work and comment on future perspectives.

%%%%%%%%%%%%%%%%%%%%%%%%%%%%%%%%%%%%%%%%%%%%%%%%%%
\section{Numerical setup}
\label{sec:model}

\subsection{Reference model}
In the following, we introduce and briefly describe the input model employed in this work, which corresponds to the fiducial case presented in P23.
The post-merger environment surrounding the central remnant through which the incipient jet has to propagate was directly imported in terms of rest-mass density, pressure, velocity, and magnetic field from the outcome of a general-relativistic magnetohydrodynamic simulation of a BNS merger presented in \cite{Ciolfi2020a}.
This realistic magnetised environment was interpolated onto a new grid in 3D spherical coordinates in order to be evolved with the special-relativistic magnetohydrodynamic module of the PLUTO code \citep{Mignone2007-PLUTO1}. 
A 90$^\circ$ axis rotation was also applied so that the new polar axis was orthogonal to the orbital axis of the merging BNS (along which the jet was then injected), avoiding any effect of the polar axis on the jet evolution.
An excision radius of 380\,km was applied in the new domain to cut out the central region so that general relativity effects that are not accounted for in PLUTO (and are not relevant for the jet evolution at large scales) can be safely neglected.
Logarithmic spacing in the radial direction was imposed to obtain the highest resolution closer to the inner boundary. The outer radial boundary was set to $r_\mathrm{max}$ = $2.5 \times 10^6$ km. Grid spacing along $r,\theta,\phi$, at $r\!=\!380$\,km, was $\approx4.4,4.4$, and $4.7$\,km, respectively, corresponding to a resolution that is at least as high as in the original merger simulation (at that distance from the origin).
Moreover, the uniform density and pressure characterising the artificial atmosphere in the merger simulation was redefined into a variable floor scaling as $r^{-6.5}$.

 For the time evolution, a piecewise parabolic reconstruction, the HLL Riemann solver, and the third order Runge-Kutta time stepping were adopted. To handle the divergence-free condition of the magnetic field, the Hyperbolic Divergence Cleaning technique was employed (see e.g. \citealt{MT2010}). Furthermore, we used the Taub-Matthews equation of state (an analytical approximation of the one derived by Synge in 1957), implemented in the PLUTO code (see \citealt{Mignone2005, Mignone2007}).
We also note that the (Newtonian) gravitational pull from the central object was included, using the mass of the BNS merger remnant, namely 2.596 M$_\odot$.

After the data import, the post-merger environment was evolved in PLUTO employing radial boundary conditions at the excision radius that were set to reproduce the angle-averaged time evolution of the different physical quantities as directly extrapolated from the BNS merger simulation. A substitution method developed in P23 was used to exploit in full the data from the merger simulation up to the latest available time (~255\,ms after merger). Then, the environment was further evolved for another 100\,ms. 
At 355\,ms after merger, the merger remnant was assumed to collapse to a black hole (BH).
For a time window of 30\,ms, the effects of the collapse were accounted for by introducing a fading acceleration term in the equations of motion aimed at reproducing the gradual decrease of radial pressure gradients due to the BH formation and subsequent accretion.
Finally, at $t_{jet}$ = 385\,ms after merger, a relativistic jet was launched. In particular, a magnetised uniformly rotating axisymmetric jet was manually injected with the following parameters: half-opening angle $\Theta_\mathrm{j} = 10^\circ$, initial Lorentz factor $\Gamma_\mathrm{j} = 3$, terminal Lorentz factor $\Gamma_{\infty} = 300$, and initial two-sided luminosity of $L_{j} = 3 \times 10^{51}$ erg/s that fades away exponentially with a characteristic time scale of 0.3 s. 
The density and pressure angular profiles within the jet were computed by solving a transverse balance equation between total pressure gradient, centrifugal force, and magnetic tension.
The jet magnetic field profile is the same as what was adopted by \cite{Geng2019} \citep[see also][]{Marti2015}.
The level of magnetisation in the jet is described by the parameter 
\begin{equation}
    \frac{L_\mathrm{j}-L_{\mathrm{j},\text{HD}}}{L_\mathrm{j}}\simeq 8~\text{\%},
\end{equation}
where $L_{\mathrm{j},\text{HD}}$ corresponds to the pure hydrodynamic luminosity.

\begin{figure}
	\includegraphics[width=\columnwidth]{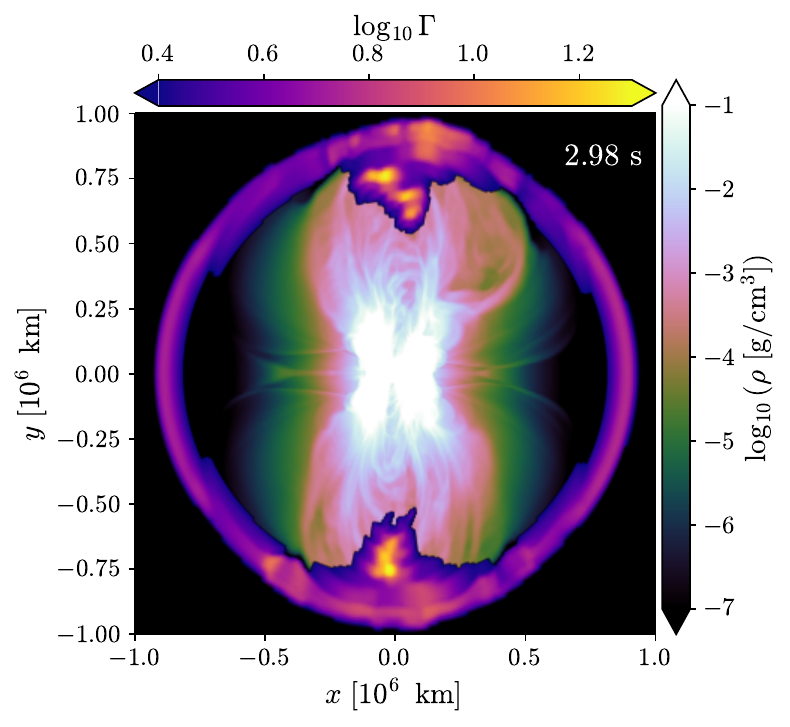}
    \caption{Meridional view of the rest-mass density and Lorentz factor 3\,s after the jet launching time, according to the fiducial simulation described in Sect.~\ref{sec:model}.}
    \label{fig:start}
\end{figure}

Within the above setup, the system is evolved for 3\,s from the jet launching time (more than the 2\,s of evolution already reported in P23). 
In Fig. \ref{fig:start}, we show a meridional view of the rest-mass density and Lorentz factor of the system at that final time, which represents the starting point for the further extension presented in the following.
For a more detailed description of the original simulation setup and results we refer to P23. 

%%%%%%%%%%%%%%%%%%%%%%%%%%%%%%%%%%%%%%%%%%%%%%%%%%
\subsection{Extended jet evolution: Towards the ballistic phase}
\label{sec:setup} 

After 3\,s of evolution, the fastest part of the outflow ($\Gamma \gtrsim 2 $) has a limited and nearly constant radial extension and the original spherical grid with a logarithmic increase in radial spacing would continuously reduce the number of radial grid points covering it.
On the other hand, a spherical grid with uniform radial spacing would increase the cell aspect ratio, causing numerical diffusion. 
Therefore, in order to maintain a 3D description with adequate
resolution in the long-term jet evolution, we switch from a spherical to a Cartesian computational grid.

Our approach consists of remapping the data at 3 s after jet launching into a new Cartesian grid, imposing that all the computational cells have cubic shape and uniform size. 
In the present work, we only follow the northern jet, but clearly the very same procedure can be applied to follow the southern one.
The Cartesian cell spacing is set to 6800\,km, which corresponds to an intermediate value between the radial spacing of the original spherical cell at the base and at the front of the high velocity region ($\Gamma \gtrsim 2 $).
With this choice, most of the jet is better resolved with respect to the original grid, except for part of the jet tail (closer to the origin). In particular, the resolution at the outer front of the jet is improved by almost a factor of two.

As a compromise between the chosen resolution and computational weight, we use $Nx = Nz = 848$ points along $x$ and $z$ and $Ny = 512$ points along $y$ (the jet injection axis), defining a computational grid with $x,z\in[-2.88, 2.88]\times 10^6$\,km, $y\in[0.1,3.58] \times 10^6$\,km. Note that a larger number of points is required in the $x,z$ directions in order to maintain a similar resolution to the $y$ direction, while working in 3D.  

In such a domain, we are able to follow the jet for additional 6.5 seconds, for a total of 9.5 seconds from $t_\mathrm{jet}$, before any of the material starts exiting the grid laterally.
The jet is evolved with the same numerical methods of the original simulation (HLL Riemann solver, parabolic reconstruction, third order Runge-Kutta time stepping, etc.). The boundary conditions (BCs) are set for every direction as ``outflow'' (zero gradient in the direction perpendicular to the boundary surface). 
We note that at the base of the domain ($y=1\times 10^5$\,km) the choice of ``outflow'' BCs along the $y$ direction leads to a persistent undesired flow of material into the domain. In fact, the residual velocity at the time the new evolution starts determines the constant value of $v_y$ enforced at all times by the BCs.
In order to limit this further injection of material (and energy), we impose an exponential decay of $v_y$ at that boundary as $v_y = v_{y,0}\,e^{-t/\tau}$ with decay time scale $\tau$ = 50 ms.
While this has no effect on the jet dynamics, it allows us to monitor the conservation of energy in the system without spurious contributions. 

%%%%%%%%%%%%%%%%%%%%%%%%%%%%%%%%%%%%%%%%%%%%%%%%%%
\subsection{Jet head definition}
\label{sec:head}

The main region of interest of the jet is where most of the energy is contained, corresponding to the part of the fluid that is more relevant for the radiative emission. 
To define the radial extension of this region, that we refer to as the jet `head', in a general way that can be applicable at any chosen time $t$ and for different simulations, we proceed as follows.

First, we remap the data back in a section of a sphere, entirely contained within the original Cartesian domain, ensuring that the largest possible grid cell is still smaller than the Cartesian grid cell, thus avoiding any loss of resolution. The interpolation is performed at $t-0.5$\,s and $t+0.5$\,s for every representative time $t$ considered.
Next, we consider a number of spherical surfaces at different radii and compute the kinetic energy flowing through each surface in one second. This is done by comparing the total kinetic energy above every surface at $t - 0.5$\,s and $t + 0.5$\,s. 
Then, we consider such variations as a representative approximation for the instantaneous time derivatives $dE/dt$, the kinetic power, at the intermediate time $t$.
\begin{figure*}
\centering
\includegraphics[width=2\columnwidth]{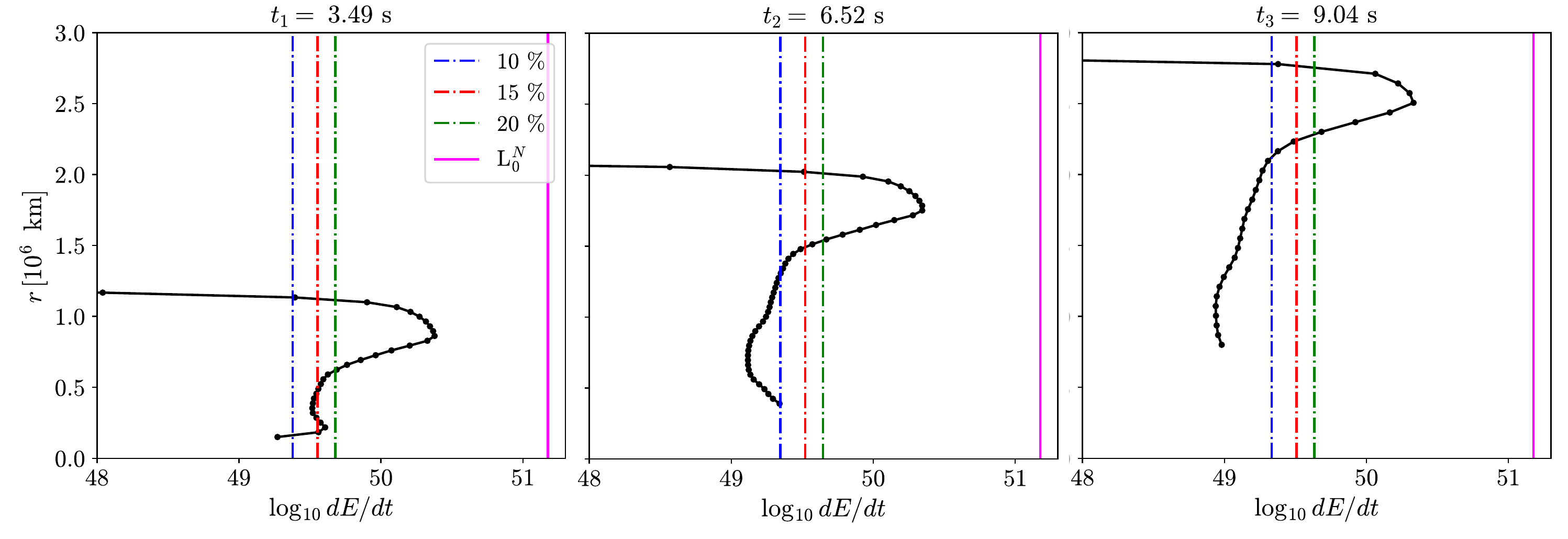}
    \caption{Kinetic power computed at different radii for the three representative times. In each panel, the blue, red, and green dot-dashed lines correspond to 10\%, 15\%, and 20\% of the maximum $dE/dt$, respectively.
    The continuous magenta line marks the initial luminosity of the jet at injection (north side only).}
    \label{fig:flux}
\end{figure*}

The result of this calculation is shown in Fig. \ref{fig:flux}, where we present $dE/dt$ as a function of the distance from the origin at three representative times (after $t_\mathrm{jet}$) of the extended simulation, namely  $t_1 \!\approx\!3.5$\,s,  $t_2  \!\approx\!6.5$\,s, and $t_3 \!\approx\!9$\,s. 
The shape of the curves reflects the time dependence of the jet luminosity at injection, with the highest power at the front and a gradual decay at smaller radii, corresponding to the fading of the injected power. From the time sequence, we can also see that the higher power portion of the jet has a nearly constant radial extension.

Moreover, Fig. \ref{fig:flux} shows (for each time or panel) three lines of constant $dE/dt$ corresponding to 10, 15, and 20\% of the maximum $dE/dt$ (blue, red, and green lines), respectively. 
We take the radial range defining the jet head at each given time as the one between the two radii where the constant $dE/dt$ line corresponding to 15\% of the maximum intersects the $dE/dt$ profile itself. 
In a later analysis we compare results obtained with thresholds at 10, 15, and 20\%, showing that the outcome is poorly sensitive to the specific choice (see Appendix \ref{sec:threshold}).
Table~\ref{tab:table} gives, for the same three times ($t_1$, $t_2$, $t_3$), the values of the corresponding minimum and maximum radii obtained by adopting either 10, 15, or 20\% as reference fraction of the maximum $dE/dt$.
\begin{table}
    \caption{Jet head definition.}
    \centering
    \begin{tabular}{c c c c c c c}
    \hline\hline
        ~ & $t_1$ & ~ & $t_2$ & ~ & $t_3$ & ~ \\ \hline
        threshold & $r_\mathrm{in}$ & $r_\mathrm{out}$ & $r_\mathrm{in}$ & $r_\mathrm{out}$ & $r_\mathrm{in}$ & $r_\mathrm{out}$ \\ \hline
        10~\text{\%}  & 0.18 & 1.13 & 1.30 & 2.02 & 2.16 & 2.78 \\ 
        15~\text{\%} & 0.49 & 1.09 & 1.51 & 1.99 & 2.30 & 2.70 \\ 
        20~\text{\%} & 0.62 & 1.09 & 1.54 & 1.99 & 2.30 & 2.70 \\ 
    \hline
    \end{tabular}
    \tablefoot{Minimum and maximum radii ($r_\mathrm{in}$ and $r_\mathrm{out}$, respectively) defining the jet head at three different times ($t_1$, $t_2$, $t_3$; see Sect.~\ref{sec:head}), based on a selection threshold of 10, 15, or 20\% of the maximum $dE/dt$ (see Sect.~\ref{sec:head} for details). Radii are given in units of $10^6$\,km.}
    \label{tab:table}
\end{table}

%%%%%%%%%%%%%%%%%%%%%%%%%%%%%%%%%%%%%%%%%%%%%%%%%%
\section{Jet evolution and energetics}
\label{sec:results}
\subsection{Jet dynamics}
\label{sec:dynamics}

\begin{figure*}
    \centering
	\includegraphics[width=1.7\columnwidth]{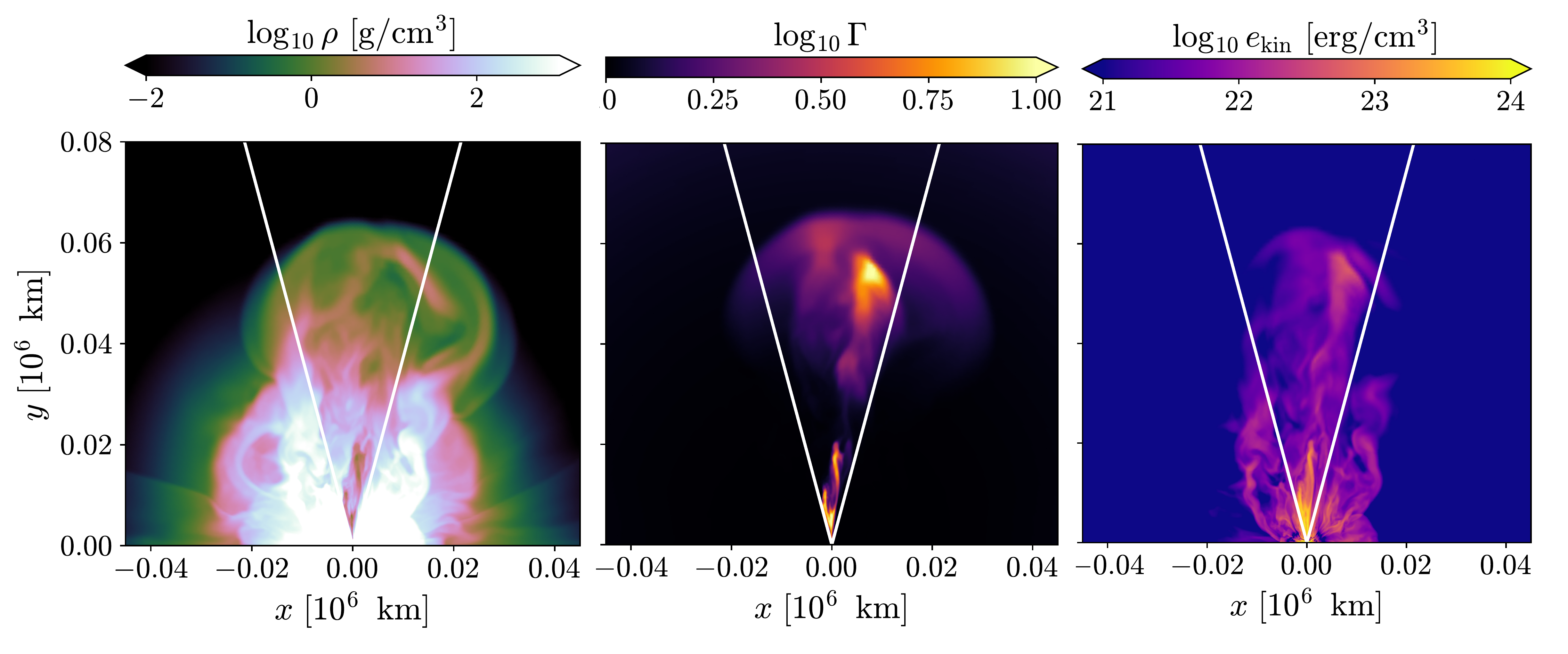}
    \caption{Meridional view of rest-mass density, Lorentz factor, and kinetic energy density (left to right) at the jet breakout time ($t_{BO}\!\simeq \!350$\,ms). The white  lines give an indicative separation between the region containing the jet and the surrounding material.}
    \label{fig:bo}
\end{figure*}

As presented in P23, complex energetic exchanges between the jet and its surrounding environment occurred during the first seconds of evolution. In the first 100 ms, the jet's movement through the dense ejecta pushes the material sideways, forming a cocoon. The breakout time, defined as the moment when the jet front first exits the post-merger environment, occurs at $t_{BO} \simeq 350$\,ms after jet launching,\footnote{In this case, we define the jet head region by considering the full outflow up to its front (located at $\sim 6.6\times 10^4$\,km).} at a distance of $r_{BO} \simeq 5 \times 10^4$ km. In Fig. \ref{fig:bo} we present different jet properties at the breakout time, obtained from the data of P23.
The realistic, non-uniform environment causes the jet front to ``snake'' through low-density regions, resulting in a non-axisymmetric shape. At this stage, the jet is still effectively collimated. In the Figure, the white lines represent indicatively its aperture (about 15$^\circ$).

\begin{figure*}
\centering
	\includegraphics[width=2\columnwidth]{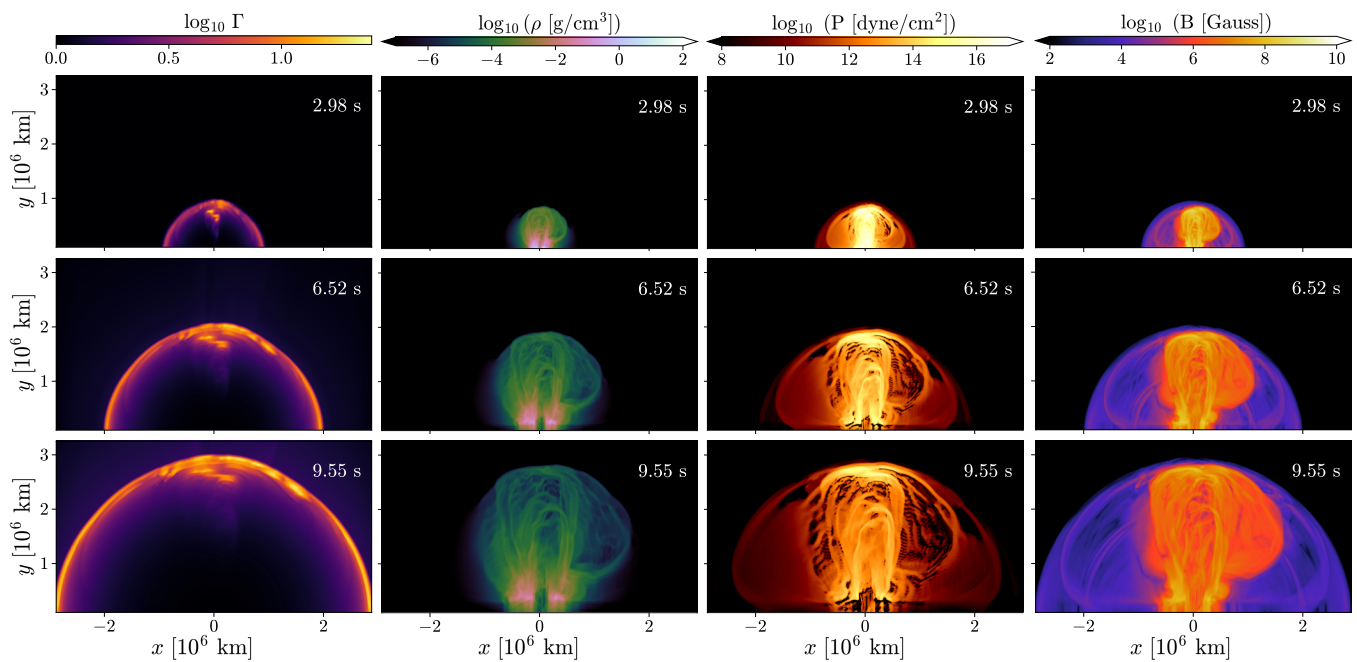}
    \caption{Meridional view of Lorentz factor, rest-mass density, total pressure, and magnetic field strength (left to right) of the simulation at three different times after jet injection (top to bottom): $ \approx\!3$\,s (beginning of the evolution on the new Cartesian grid), $\approx\!6.5$\,s, and $\approx\!9.5$\,s.}
    \label{fig:cart_panel}
\end{figure*}

In Fig. \ref{fig:cart_panel}, we present some properties of the jet at different times of evolution in the new Cartesian grid. Lorentz factor, rest-mass density, pressure, and magnetic field strength are shown on the $xy$-plane at the initial, intermediate, and final stages of evolution: at  about 3, 6.5, and 9.5\,s after $t_\mathrm{jet}$, respectively.
From the distribution of the Lorentz factor, we observe that the high velocity part of the outflow consists of the radially expanding jet front followed by a short tail surrounded by a spherical shell accelerated at earlier times, which is of extremely low density and energetically unimportant (see below).
The observed radial stratification in the tail is associated to the fading behaviour of the injected luminosity.
During the evolution, there is no significant radial spread of the high velocity region ($\Gamma \gtrsim 2 $), while the jet continues to expand laterally, becoming more and more similar to a portion of a spherical shell with a nearly constant half-opening angle of about 30$^\circ$. 
At this stage, the jet is no longer interacting with the environment surrounding the merger site and only undergoes further acceleration and slight morphological evolution while expanding across the artificial floor medium, which has negligible impact on the dynamics. 

In the first seconds of evolution, the jet entrains a large amount of environmental material (see Fig. \ref{fig:start}) that is only in part imported in the new Cartesian domain.
During our simulation the rest-mass density distribution, shown in the second column of Fig. \ref{fig:cart_panel}, indicates that such material is expanding slowly, with a low density funnel previously excavated by the jet along the $y$-axis.
At the lower boundary ($y=10^5$\,km), we see the effects of the exponentially decaying velocity $v_y$ imposed to damp further injection of matter and energy, as discussed in the previous Section \ref{sec:setup}. Analogous effects can be seen in the pressure and magnetic field strength (third and fourth columns). 

\begin{figure}
	\includegraphics[width=0.9\columnwidth]{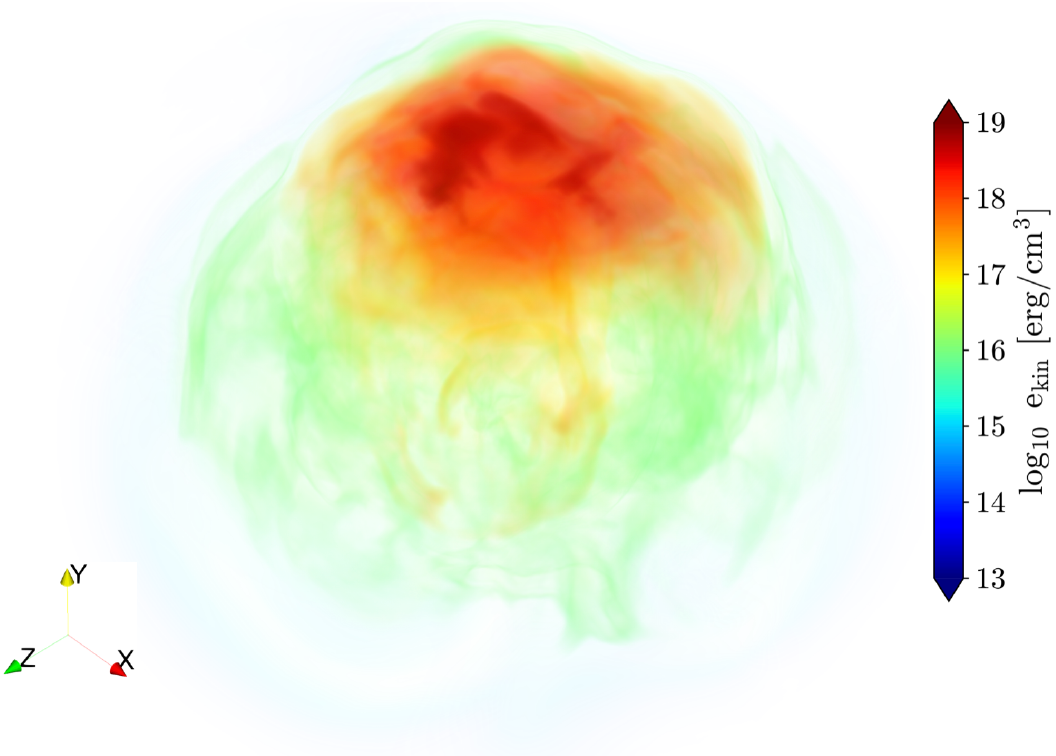}
    \caption{Three-dimensional rendering of the kinetic energy density at the final simulation time ($\!\approx\!9.5$\,s after jet injection). Higher transparency is applied to the lowest values of energy to allow for a better visualisation. Spatial scale is limited to $y\in [ 1,3.5 ]\times 10^6$\,km.}
    \label{fig:3d}
\end{figure}

To get a better visualisation of the distribution of the outflow material in the whole domain, a three-dimensional plot of the kinetic energy density at the end of the simulation is reported in Fig.~\ref{fig:3d}. It is evident that the energy distribution is highly non-axisymmetric (see also Sect. \ref{sec:angular}) and the contribution of the shell observed in the Lorentz factor (see Figs.~\ref{fig:start} and \ref{fig:cart_panel}) is negligible at large angles ($\gtrsim\!30^\circ$) from the jet injection axis. 

%%%%%%%%%%%%%%%%%%%%%%%%%%%%%%%%%%%%%%%%%%%%%%%%%%
\subsection{Energy budget and ballistic regime}
\label{sec:energy}

\begin{figure}
\centering
	\includegraphics[width=\columnwidth]{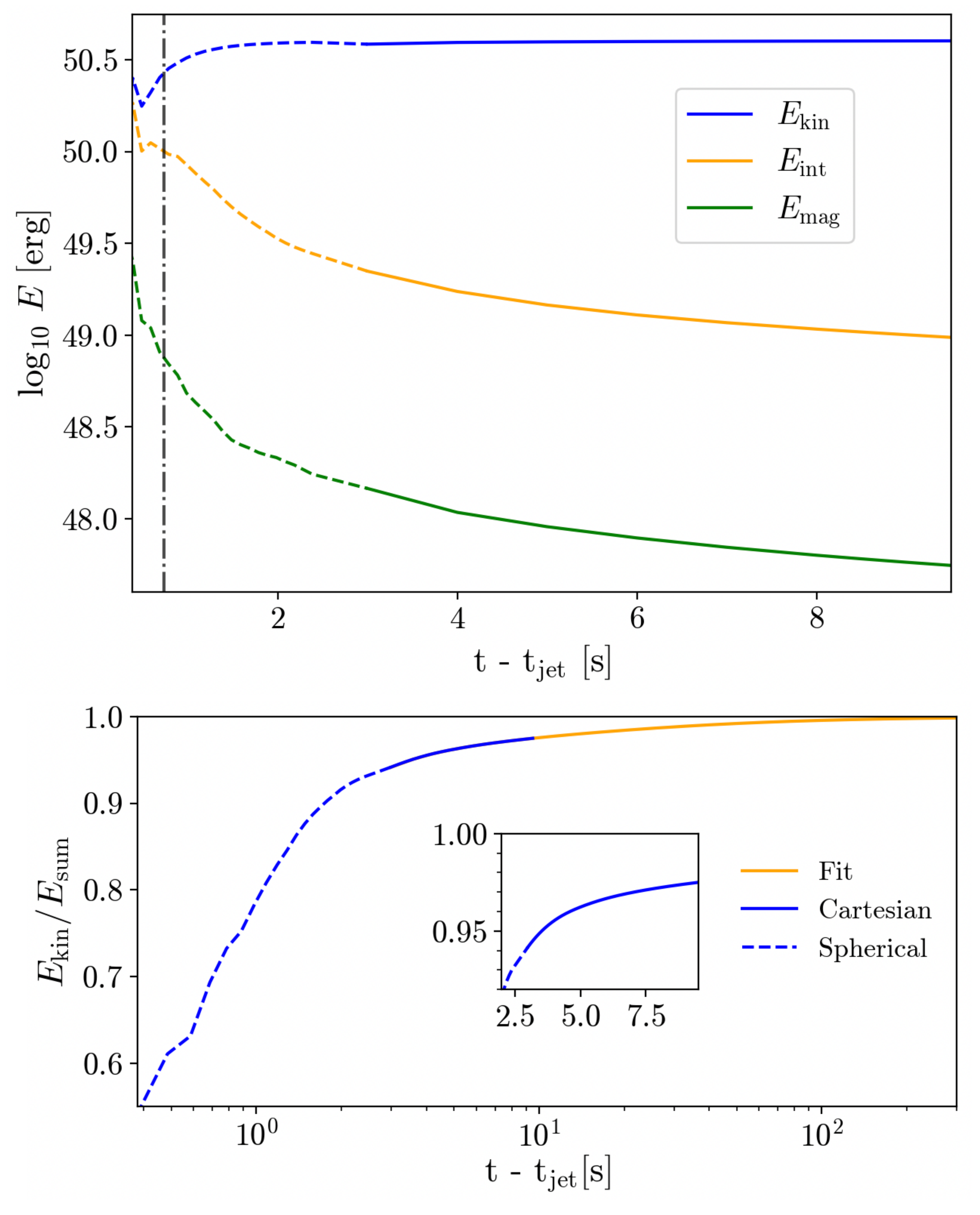}
    \caption{Time evolution of the different energy components in the whole computational domain. Upper panel: kinetic ($E_\mathrm{kin}$), internal ($E_\mathrm{int}$), and magnetic ($E_\mathrm{mag}$) energies are represented, with dashed (continuous) lines referring to the evolution employing the spherical (Cartesian) grid. The black vertical line represents the breakout time (see text). Lower panel: Evolution of the fraction ${E}_\mathrm{kin}/{E}_\mathrm{sum}$ (see text) within the simulation time in the original spherical grid (dashed blue line) and after remapping into the new Cartesian grid (full blue line), further extrapolated to later times via a polynomial fit (orange line). Time is represented in logarithmic scale. In the inset, we show with a linear scale in time the behaviour of the same energy ratio referred only to the extended Cartesian simulation.}
    \label{fig:ecomp}
\end{figure}

In this Section, we discuss the evolution of the jet energetics with time, including also the data for the first seconds of evolution from the reference simulation.
The upper panel of Fig. \ref{fig:ecomp} shows the evolution of the kinetic ($E_\mathrm{kin}$), internal ($E_\mathrm{int}$), and magnetic ($E_\mathrm{mag}$) energies over the whole domain.
In the very first milliseconds of evolution all the energetic components decrease, due to the effects of the gravitational attraction from the inner radial boundary (see P23). 
Subsequently, internal and magnetic components are continuously converted into kinetic form.

Focusing the attention on the extended evolution ($t\!>\!t_\mathrm{jet}+3$\,s), we observe that the energy $E_\mathrm{sum}\equiv E_\mathrm{kin}+E_\mathrm{int}+E_\mathrm{mag}$ in the new Cartesian domain is approximately $4.1 \times 10^{50}$ erg, very well conserved throughout the extended simulation. At the beginning, energy conservation is slightly altered by the spurious injection of material at the lower $y$ boundary. However, this effect is removed after the first second due to the exponential decay imposed on $v_y$ at that boundary (see above). Quantitatively, $E_\mathrm{sum}$ increases by about 0.7\% in the first second (from 3 to 4\,s after jet launching) and by only 0.08\% from 4\,s to the end. Such a good level of conservation shows that the increase of kinetic energy corresponds precisely to the decrease in magnetic and internal energies, and therefore no significant errors are present in the numerical simulation nor in the way the energies were computed over the computational domain.

At 3 s after $t_\mathrm{jet}$, when the Cartesian simulation begins, about 94\% of $E_\mathrm{sum}$, namely ~the energy that is available to be converted into radiation at later times, is kinetic. 
The lower panel of Fig.~\ref{fig:ecomp} shows how the fraction of kinetic energy keeps increasing up to about 98\% at the final time of the simulation.
In the Figure, a polynomial extrapolation of the increasing trend of the kinetic fraction is also shown to give an idea of how long it would take to reach a 99.9\% conversion, corresponding to more than 200\,s (see also \citealt{Xie2018}).

Based on these results, 
the assumption that the jet has reached a ballistic expansion phase with full energy conversion at the end of our simulation would imply an error of only about 2\% in the jet kinetic energy.
This also translates in Lorentz factors of the fluid elements that are, on average, lower than the one at saturation by approximately a factor $\sqrt{E_\mathrm{sum}/E_\mathrm{kin}}=\sqrt{1/0.975}$: the difference is of only 1\%. We refer to Sect. \ref{sec:saturation} for a more detailed analysis on the saturation limit.

%%%%%%%%%%%%%%%%%%%%%%%%%%%%%%%%%%%%%%%%%%%%%%%%%%
\section{Jet structure}
\label{sec:angular}

Here, we analyse the main properties of the most energetic part of the jet, quantitatively defined as `head' (see \ref{sec:head}). Specifically, we focus on two of the relevant quantities for afterglow emission, namely, the radial and angular distributions of energy and Lorentz factor.
Hereafter, the relevant data are remapped in spherical coordinates ($r,\theta,\phi$), as described in \ref{sec:head}.

%%%%%%%%%%%%%%%%%%%%%%%%%%%%%%%%%%%%%%%%%%%%%%%%%%
\subsection{Energy distribution}
\label{sec:en_distr}

The afterglow signal depends on the jet radial and angular kinetic energy distributions, once these are essentially frozen.
We examine how such energy is distributed towards the end of the simulation, while 
the evolution (and freezing) of the angular distributions of Lorentz factor and energy per unit solid angle is addressed in the following subsections.  

\begin{figure}
\centering
	\includegraphics[width=0.92\columnwidth]{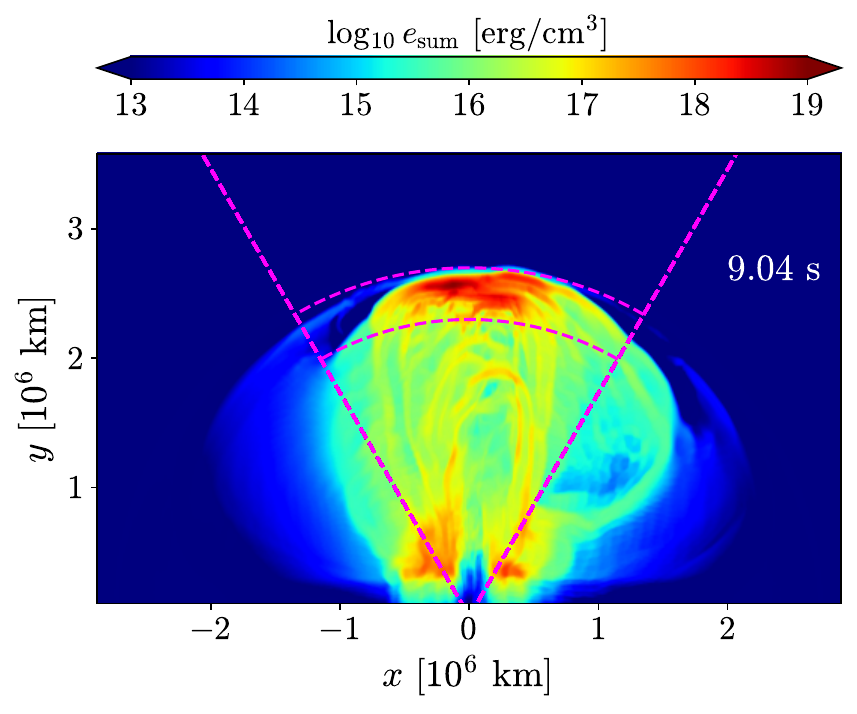}
    \caption{Meridional view of the energy density $e_\mathrm{sum}$ at $\!\approx\!$ 9\,s after the jet launching time. The dashed magenta lines mark (i) the radial range of the jet `head' (according to the definition given in Sect. \ref{sec:head}), and (ii) the region within 30$^\circ$ half-opening angle from the jet injection axis.}
    \label{fig:enang}
\end{figure}

In Fig.~\ref{fig:enang}, we show again a meridional view of the energy density $e_\mathrm{sum}$ at $t_3$, where we now mark the region within 30$^\circ$ from the jet injection axis (see below) and, inside such opening angle, the radial range defining the jet head at this time of the evolution. 
We also note that, at the given time, the jet head, which by definition contains the most energetic part of the jet, has evolved into a radially thin portion of a spherical shell.

\begin{figure}
\centering
    \includegraphics[width=0.92\columnwidth]{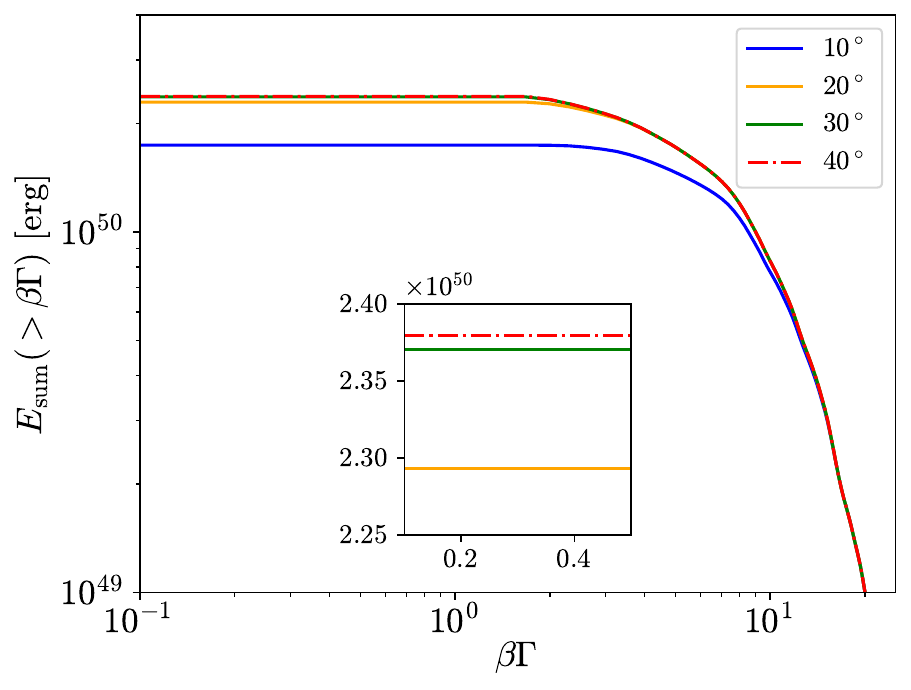}
    \caption{Total energy $E_\mathrm{sum}$ (without rest-mass energy contribution) of the fluid elements within the jet head having $\beta\Gamma$ larger than a certain value, as a function of such a value. The different lines and colours correspond to the result obtained by limiting the calculation within 10$^\circ$, 20$^\circ$, 30$^\circ$, and 40$^\circ$ from the jet injection axis. 
    The plot refers to $\!\approx\!$ 9\,s after the jet launching time.
    The inset shows a selected zoomed-in portion of the same profiles. 
    }
    \label{fig:gmbet} 
\end{figure}

Figure~\ref{fig:gmbet} illustrates how the energy content of the jet head is distributed in terms of faster and slower outflow components. Namely, we plot $E_\mathrm{sum}$ relative only to the fluid elements with $\beta\Gamma$ larger than a certain value, as a function of such a value, hereafter $E_\mathrm{sum}(>\beta\Gamma)$. 
Moreover, we consider only portions of the jet within a given angular distance from the jet injection axis, comparing the results obtained for different choices of the half-opening angle, namely, 10$^\circ$, 20$^\circ$, 30$^\circ$, and 40$^\circ$.
The profiles for 10$^\circ$ and 20$^\circ$ differ by about 30\%, while those for 20$^\circ$ and 30$^\circ$ show a discrepancy of about an order of magnitude less, and for 30$^\circ$ and 40$^\circ$ the profiles differ by $\lesssim1$\%. 
We conclude that most of the jet energy is comprised within 30$^\circ$ from the injection axis.

We stress that most of the jet energy is in fluid elements with $\beta\Gamma \gtrsim 3$.
Furthermore, the faster portion is fully contained within a half-opening angle $<10^\circ$, as shown by the absence of discrepancies between the different profiles for $\beta\Gamma \gtrsim 10$.

We note that the Lorentz factor of the energetically relevant parts of the jet does not exceed $\sim$20, which is considerably lower than the theoretical maximum set at jet launching, $\Gamma_\infty\!=\!300$. This is due to the strong energy losses that occur while the jet is making its way out of the environment.
It would be interesting to estimate how much of the injected jet energy is transferred to the surrounding during its propagation ($E_\mathrm{surr}$). However, this estimate is not at all straightforward since we cannot easily determine how the central accretion possibly affects the energy budget.  Therefore, we are only able to set upper limits to $E_\mathrm{surr}$, by assuming that the accretion is only affecting the post-merger environment.  Another issue is to properly distinguish the jet and the surrounding material. We consider two extremes for the estimate of the jet energy:  the jet head region, and the whole volume contained within 30$^\circ$. The former provides a lower limit, while the latter results in an upper limit since it is contaminated by the energy of the post-merger environment.
By comparing the energy injected in the first 3\,s, $E_\mathrm{sum}^\mathrm{inj}(3\,\mathrm{s}) \simeq 4.63 \times 10^{50}$\,erg, with the energy of the jet head at the same time, $E_\mathrm{sum}^\mathrm{head}(3\,\mathrm{s}) \simeq 2.59\times 10^{50}$\,erg, we find that at most 44\% of energy is potentially transferred to the surroundings. While by comparing $E_\mathrm{sum}(3\,\mathrm{s}, <30^\circ) \simeq 3.65 \times 10^{50}$\,erg with the injected energy, we find an upper limit $E_\mathrm{surr}$ < 21\% of $E^\mathrm{inj}_\mathrm{sum}$. 
These upper limits are consistent with the fact that the jet achieves a rather limited maximum Lorentz factor at the end of the simulation. 

It is clear that the specific model we are considering does not represent a jet that could account for a typical sGRB. Nonetheless, it is still expected to produce radiation at larger scales, that will be analysed in a future work.
Moreover, we remark that the methods presented here are fully applicable to a wide range of physical cases.
\begin{figure*}
\centering
	\includegraphics[width=2.\columnwidth]{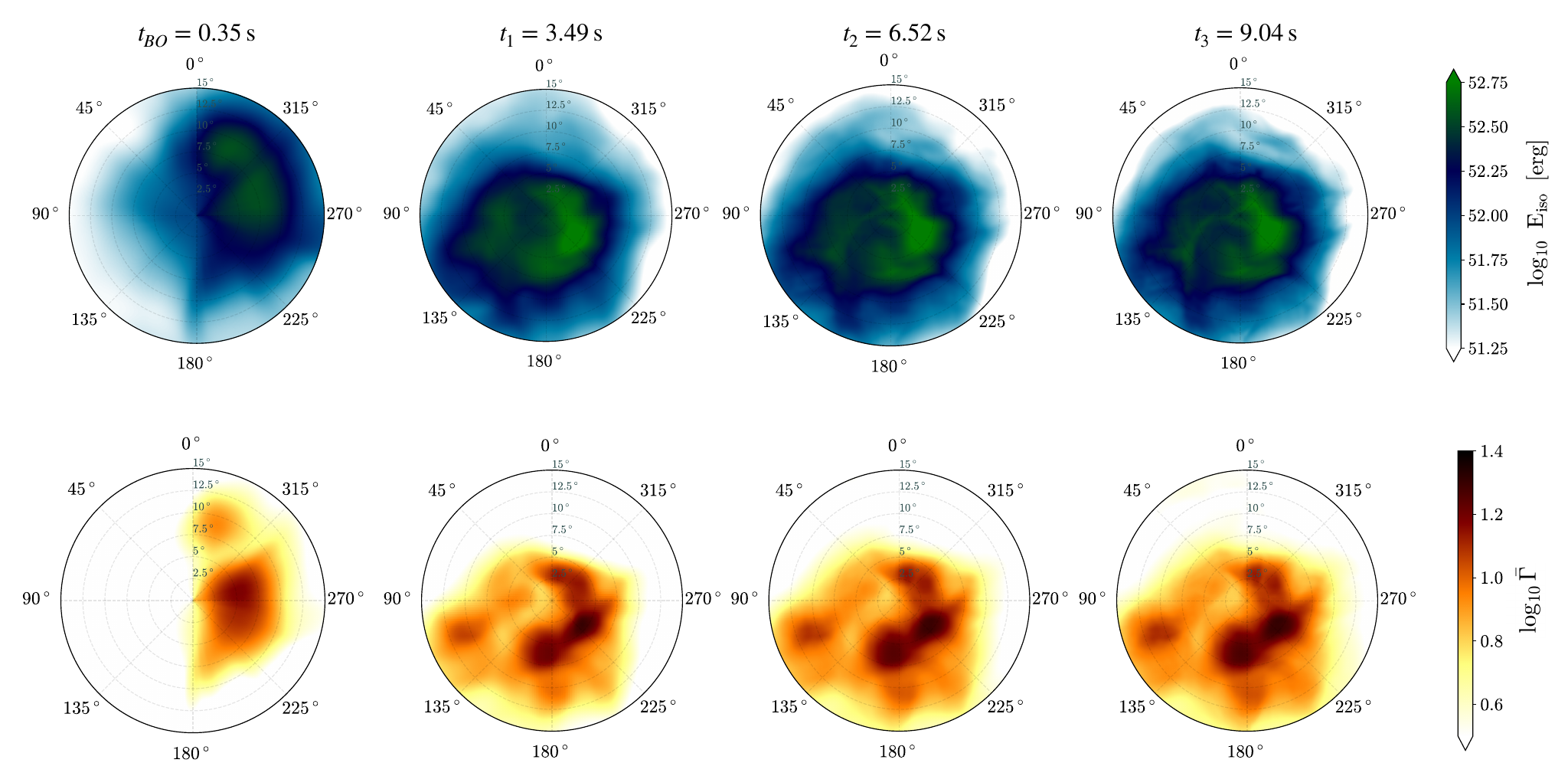}
    \caption{Two-dimensional angular distribution of the isotropic equivalent energy (top) and the radially averaged Lorentz factor (bottom) of the jet head at times $t_{BO}$, $t_1$, $t_2$ and $t_3$. The jet injection axis corresponds to the central point in each panel.
    }
    \label{fig:egm}
\end{figure*}

%%%%%%%%%%%%%%%%%%%%%%%%%%%%%%%%%%%%%%%%%%%%%%%%%%
\subsection{Front-view angular distributions}
\label{sec:ang2d}

In this Section, we discuss the angular distribution of Lorentz factor and energy per unit solid angle characterising the jet head.
In order to have a representative value of the Lorentz factor for each angular position with respect to the jet injection axis, we define an energy-weighted radial average $\bar{\Gamma}$ as follows:
\begin{equation}
    \bar{\Gamma}(\Theta,\Phi) = \frac{\int_{r_\mathrm{in}}^{r_\mathrm{out}} \Gamma(r,\Theta,\Phi) e_{\mathrm{sum}}(r,\Theta,\Phi) r^2  dr}{\int_{r_\mathrm{in}}^{r_\mathrm{out}}  e_{\mathrm{sum}}(r,\Theta,\Phi) r^2 dr}  \,\, ,
\label{gammabar}
\end{equation}
where $(\Theta,\Phi)$ indicate the polar and azimuthal angles with respect to the jet injection axis, (not to be confused with the actual polar coordinates $(\theta,\phi)$, defined with respect to a polar axis orthogonal to the injection axis, see Sect.~\ref{sec:model}).
The choice of weighting over $e_\mathrm{sum}$ is motivated by the fact that this represents the available energy that could be converted into radiation. 

Next, we consider the distribution of the isotropic equivalent energy. In our case it is defined as
\begin{equation}    
    E_{\mathrm{iso}}(\Theta,\Phi) = 4 \pi \frac{dE_\mathrm{sum}}{d\Omega} = 4 \pi \int_{r_\mathrm{in}}^{r_\mathrm{out}} r^2 e_{\mathrm{sum}}(r,\Theta,\Phi) dr \,\, ,
\end{equation}
where $dE/d\Omega$ is the energy per unit solid angle along each $(\Theta,\Phi)$ direction.

The 2D front-view angular distributions (in $\Theta,\Phi$) of $\bar{\Gamma}$ and $E_{\mathrm{iso}}$ are shown in Fig. \ref{fig:egm} for the breakout time $t_{BO}$ and the three representative times $t_1, t_2, t_3$.
A first important result is that the angular distributions are not axisymmetric.
The maximum values of both the energy and Lorentz factor are about 4$^\circ$ away from the jet injection axis. By comparing the profiles at the breakout time with the ones obtained from the extended simulation, we observe an evolution of the jet structure, that evolves from what seems to be compatible with a tilted jet, to having the fastest part of the outflow distributed in a sort of annular shape around the axis. Such a feature is reminiscent of the `hollow core' pointed out in previous GRB jet simulations (e.g. \citealt{Nathanail2021} and refs.~therein). This could be related to the structure of the injected magnetic field, having the maximum of the toroidal component at 4$^\circ$ (see \citealt{Geng2019}), together with the overall tilting of the entire jet due to the strong interaction with the environment.
Moreover, we note a fragmentation of the fastest portion of the jet into a number of local maxima of Lorentz factor, which can also be expected in cases where the jet has a hard time piercing through a dense and non-uniform medium.
These aspects will be further analysed in a future work, where a number of jet simulations with different injection parameters will be compared.

The overall shape of the jet remains mostly unchanged throughout the Cartesian simulation, consistently with the fact that the jet is not anymore interacting with the post-merger environment.
However, as shown by the quantitative analysis presented in the next Section and further discussed in Sect. \ref{sec:saturation}, there are still differences between $t_1$ and $t_3$ that are significant for predicting the afterglow emission.

%%%%%%%%%%%%%%%%%%%%%%%%%%%%%%%%%%%%%%%%%%%%%%%%%%
\subsection{Angular profiles}
\label{sec:ang1d}

\begin{figure*}
\centering
	\includegraphics[width=2\columnwidth]{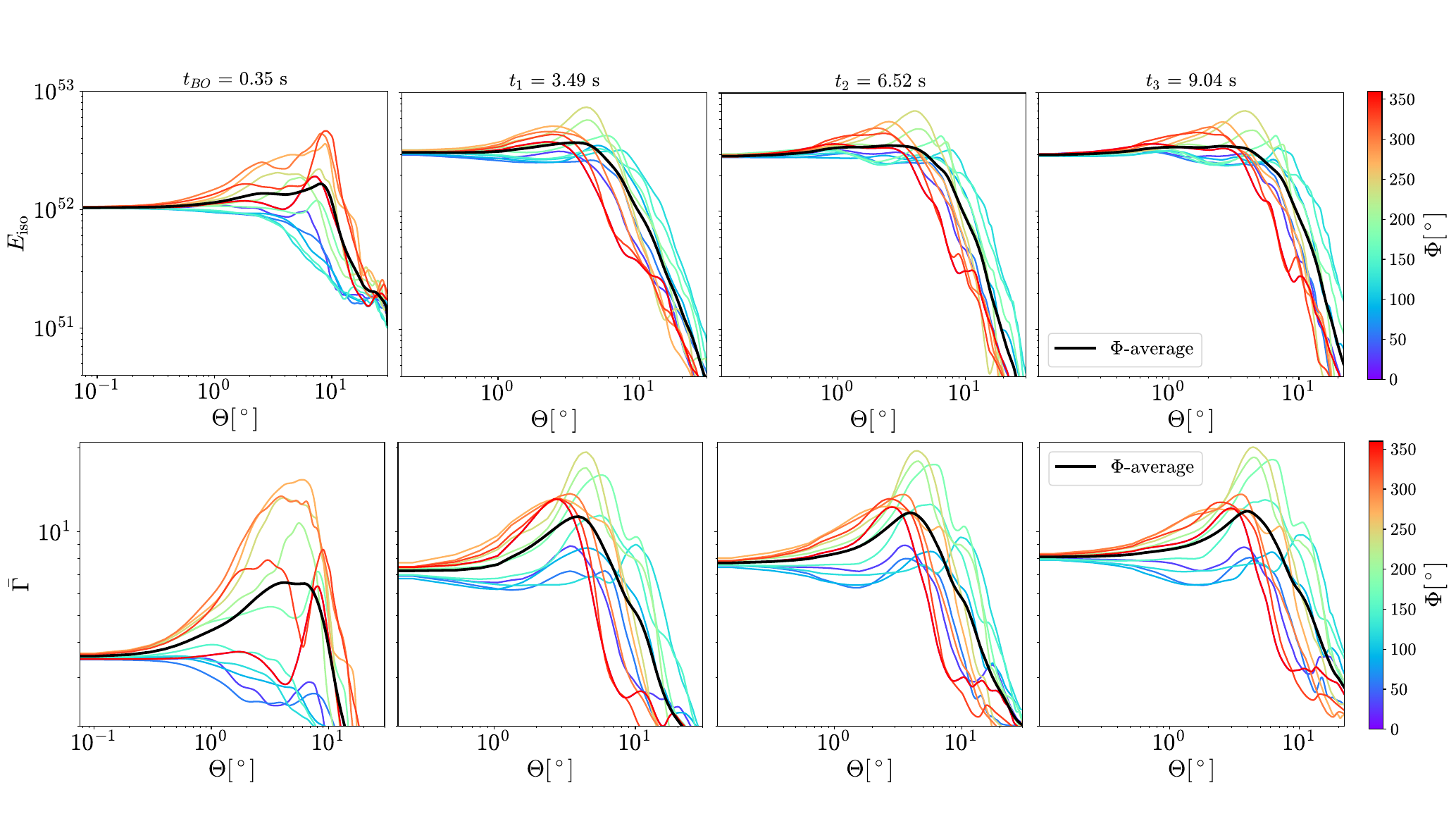}
    \caption{Isotropic equivalent energy (top) and radially averaged Lorentz factor (bottom) of the jet head (see Sect.~\ref{sec:head}) as functions of the angular distance from the jet injection axis $\Theta$, at four different times (after jet injection).
    Different colours correspond to different azimuthal angles $\Phi$ and the thick blue line represents the $\Phi$-averaged results. All quantities are in logarithmic scale.
    }
    \label{fig:1d}
\end{figure*}

In the following analysis, the angular profiles of $\bar{\Gamma}(\Theta,\Phi)$ and $E_\mathrm{iso}(\Theta,\Phi)$ (see Sect. \ref{sec:ang2d}) are examined as functions of $\Theta$ alone for each selected value of the jet azimuthal angle $\Phi$.\footnote{The impact of the spherical-to-Cartesian and Cartesian-to-spherical interpolations (Sect.~\ref{sec:setup} and \ref{sec:head}) on the accuracy of such angular profiles is addressed in Appendix \ref{sec:interpolation}.}
The resulting profiles are presented in Fig. \ref{fig:1d}, where in each panel a number of lines referring to different values of $\Phi$ are shown together with the profile obtained by averaging over $\Phi$ (thick line), namely:
\begin{equation}
    \bar{\Gamma}(\Theta)|_{\Phi_\mathrm{avg}} = \frac{\int_0^{2\pi} \bar{\Gamma}(\Theta,\Phi) d\Phi}{2\pi} \, ,
\end{equation}
\begin{equation}
    E_{\mathrm{iso}}(\Theta)|_{\Phi_\mathrm{avg}} = \frac{\int_0^{2\pi} E_\mathrm{iso}(\Theta,\Phi) d\Phi}{2\pi}  \, .
\end{equation}

For all the represented profiles of both quantities, we observe an almost flat core within $\Theta \lesssim 1^\circ$, a peak comprised between $2^\circ \lesssim \Theta \lesssim 10^\circ$, and a decaying trend at larger angles corresponding to the lateral `wings' of the outflow. The curves representing different azimuthal angles show a rather large dispersion, with a factor $\approx\!3$ variation in both energy and Lorentz factor around $\Theta\simeq4^\circ$ (particularly evident for $\bar{\Gamma}$, for which the maximum value is up to 70\% larger than the value along the axis) and up to one order of magnitude at larger angles. 
Such a scatter implies that modelling the angular distributions and consequently the jet afterglow emission under the assumption of axisymmetry can potentially lead to significant 
errors. 
This affects any estimate based on the comparison with afterglow observations where axisymmetry is assumed (e.g. the case of GRB\,170817A). 

\begin{figure*}
\centering
    \includegraphics[width=1.96\columnwidth]{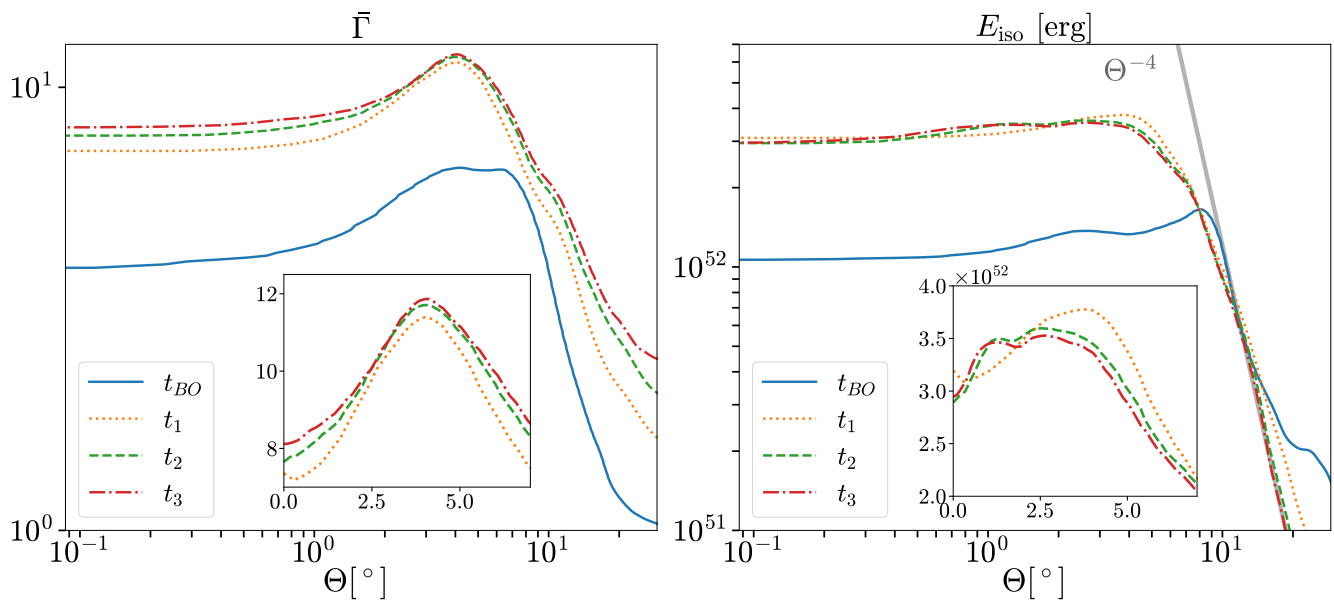}
    \caption{$\Phi$-averaged profiles (see Fig.~\ref{fig:1d}) of the radially averaged Lorentz factor (left) and isotropic equivalent energy (right) of the jet head as functions of $\Theta$. In order to highlight how the angular distributions evolve with time, in each panel we plot together the result for four different times.}
    \label{fig:angcfr}
\end{figure*}

In Fig. \ref{fig:angcfr}, we focus the attention on the time evolution of the angular profiles in terms of the $\Phi$-averaged ones.
The comparison at different times
confirms the trend identified with the 2D angular distributions, specifically, the maximum located 4$^\circ$ away from the jet injection axis (see Fig. \ref{fig:egm}), and an evident acceleration and evolving energy distribution after $t_{BO}$. Moreover, we still observe changes from $t_1$ to $t_2$: 
$\bar{\Gamma}$ maintains a similar profile but grows in value, while the $E_\mathrm{iso}$ profile tends to flatten in the range $(1-4)^\circ$ and the wings at $\gtrsim\!10^\circ$ evolve into a steeper slope $\propto \Theta ^{-4}$.
The differences between $t_2$ and $t_3$ are strongly reduced,
confirming that around $t_3$ the angular distributions are nearly frozen. 
Towards the end of the simulation, the maximum $\bar{\Gamma}(\Theta)|_{\Phi_\mathrm{avg}}$ is $\simeq12$ and the maximum $E_\mathrm{iso}(\Theta)|_{\Phi_\mathrm{avg}}$ is $\simeq3.5\times10^{52}$\,erg.

%%%%%%%%%%%%%%%%%%%%%%%%%%%%%%%%%%%%%%%%%%%%%%%%%%
\section{Towards an afterglow emission modelling}
\label{towards-afterglow}

In this Section, we discuss two further quantities that are relevant for calculating the afterglow light curves from jet simulations like the one presented in this work: the angular profile of the saturation Lorentz factor and
the scale distance at which the afterglow emission is expected to be produced.

%%%%%%%%%%%%%%%%%%%%%%%%%%%%%%%%%%%%%%%%%%%%%%%%%%
\subsection{Saturation limit}
\label{sec:saturation}

Here, we want to use the obtained profile of the radially and azimuthally averaged Lorentz factor of the jet head to estimate the profile of the saturation (or asymptotic) value of $\Bar{\Gamma}_s$, that would be reached when the conversion of energy in kinetic form is complete.
Such $\Bar{\Gamma}_s$ is the appropriate quantity to be employed as input for computing the afterglow emission.
In the following, we discuss how we obtain such an estimate and how the result changes from 3\,s to 9\,s after jet launching. 

The lab-frame kinetic energy density of a representative element of the outflow is given by
\begin{equation*}
    e_\mathrm{kin} = \rho\Gamma(\Gamma-1)c^2 \, ,
\end{equation*}
where $\Gamma$ is its Lorentz factor and $\rho$ the rest-mass density of the element in the comoving frame. We can then estimate a saturation limit by substituting $e_\mathrm{kin}$ with $e_\mathrm{sum}$, thus assuming that all the energy has been converted into kinetic, namely:
\begin{equation*}
    \Gamma_s \approx \sqrt{e_\mathrm{sum}/e_\mathrm{kin}} \, \Gamma \, ,
\end{equation*}
where we approximated $\Gamma(\Gamma-1)\approx \Gamma^2$ and $\Gamma_s(\Gamma_s-1)\approx \Gamma_s^2$.

We thus computed the saturation radially averaged Lorentz factor of the jet head $\bar{\Gamma}_s$ as (see Eq.~(\ref{gammabar}))
\begin{equation}
    \bar{\Gamma}_{s}(\Theta,\Phi) =  \frac{\int_{r_\mathrm{in}}^{r_\mathrm{out}}  \Gamma  e_\mathrm{sum} (r,\Theta,\Phi) \sqrt{e_\mathrm{sum}(r,\Theta,\Phi)/e_\mathrm{kin}(r,\Theta,\Phi)} \,  r^2 dr} {\int_{r_\mathrm{in}}^{r_\mathrm{out}} e_\mathrm{sum}(r,\Theta,\Phi) r^2 dr} \, ,
\end{equation}
and then obtained the $\Phi$-averaged profile $\bar{\Gamma}_s(\Theta)$.
\begin{figure}
	\includegraphics[width=0.91\columnwidth]{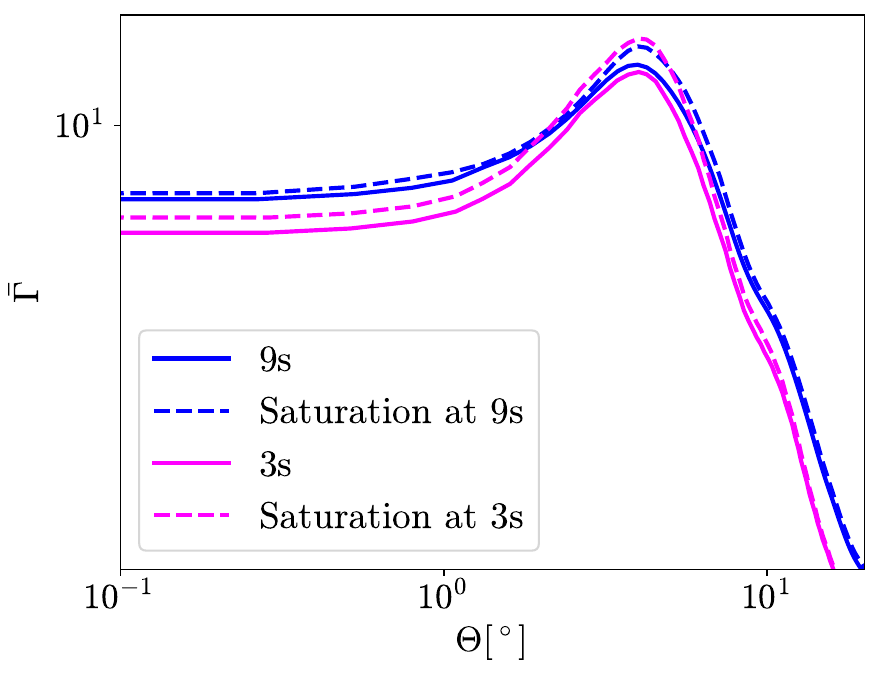}
    \caption{Angular profile of the radially and azimuthally averaged Lorentz factor of the jet head at 3\,s and 9\,s (continuous magenta and blue lines, respectively), along with the corresponding saturation profiles (dashed lines; see text).} 
    \label{fig:satr}
\end{figure}

Figure~\ref{fig:satr} shows together the original and saturation profiles referring to 3 and 9\,s.
A first observation is that the simulation profile at 9\,s is significantly closer to the corresponding saturation, with discrepancies between 2\% and 5\% up to $\Theta\!=\!10^\circ$ and smaller at larger angles, while at 3\,s the difference is up to about 10\%. 

Then, comparing the two saturation profiles at 3\,s and 9\,s, the discrepancy is only about 2\% at the peak, but it increases to 8\% (35\%) at smaller (larger) $\Theta$.
Since afterglow emission is rather sensitive to the input Lorentz factor profile, 
the overall light curve may be significantly different if predicted using the information at 3\,s or at 9\,s.
We will quantify such differences in a future work, where the afterglow emission will be investigated for a set of jet simulations, including the one considered here.
As shown by the above analysis, extending a 3D sGRB jet simulation up to several seconds or more can be useful to assess the inevitable uncertainties on the afterglow emission related to the residual jet evolution.

%%%%%%%%%%%%%%%%%%%%%%%%%%%%%%%%%%%%%%%%%%%%%%%%%%
\subsection{Deceleration radius}
\label{decel}

One of the main physical parameters required for the estimate of the expected afterglow signal is the (radial) location of the external shock at the time it starts decelerating: the so-called deceleration radius $R_\mathrm{dec}$.
The results of our simulation as discussed above grant us the availability of the three-dimensional distribution of the coasting values of the fluid velocity and energy.
With this information, we are able to provide an order-of-magnitude estimate of the scale distance $R_\mathrm{dec}$ where such a shock deceleration occurs.
 
We perform our calculations within the simplified framework of the (spherically symmetric) fireball model \citep[]{1992MNRAS.258P..41R,1993ApJ...405..278M,M_sz_ros_2006} with energy $E_\mathrm{iso}$ and bulk Lorentz factor $\Gamma$. In general, a realistic assumption for a fluid moving at relativistic velocities is the absence of lateral expansion: every fluid element at different angles evolves independently from their surrounding. 

Considering a relativistic fireball travelling in a uniform ISM with baryon number density $n$, the deceleration radius is defined by the condition that the isotropic-equivalent mass of the ISM swept-up by the jet equals its initial isotropic-equivalent energy $E_\mathrm{iso}$ \citep[]{1992MNRAS.258P..41R,Salafia2022}, namely:
\begin{equation}
    R_\mathrm{dec} = \left( \frac{3 E_\mathrm{iso}}{4 \pi m_p n c^2 \Gamma^2 }\right) ^{1/3} \, ,
    \label{eq:rd}
\end{equation}
where $m_p$ is the proton mass.

\begin{figure}
\centering
	\includegraphics[width=0.93\columnwidth]{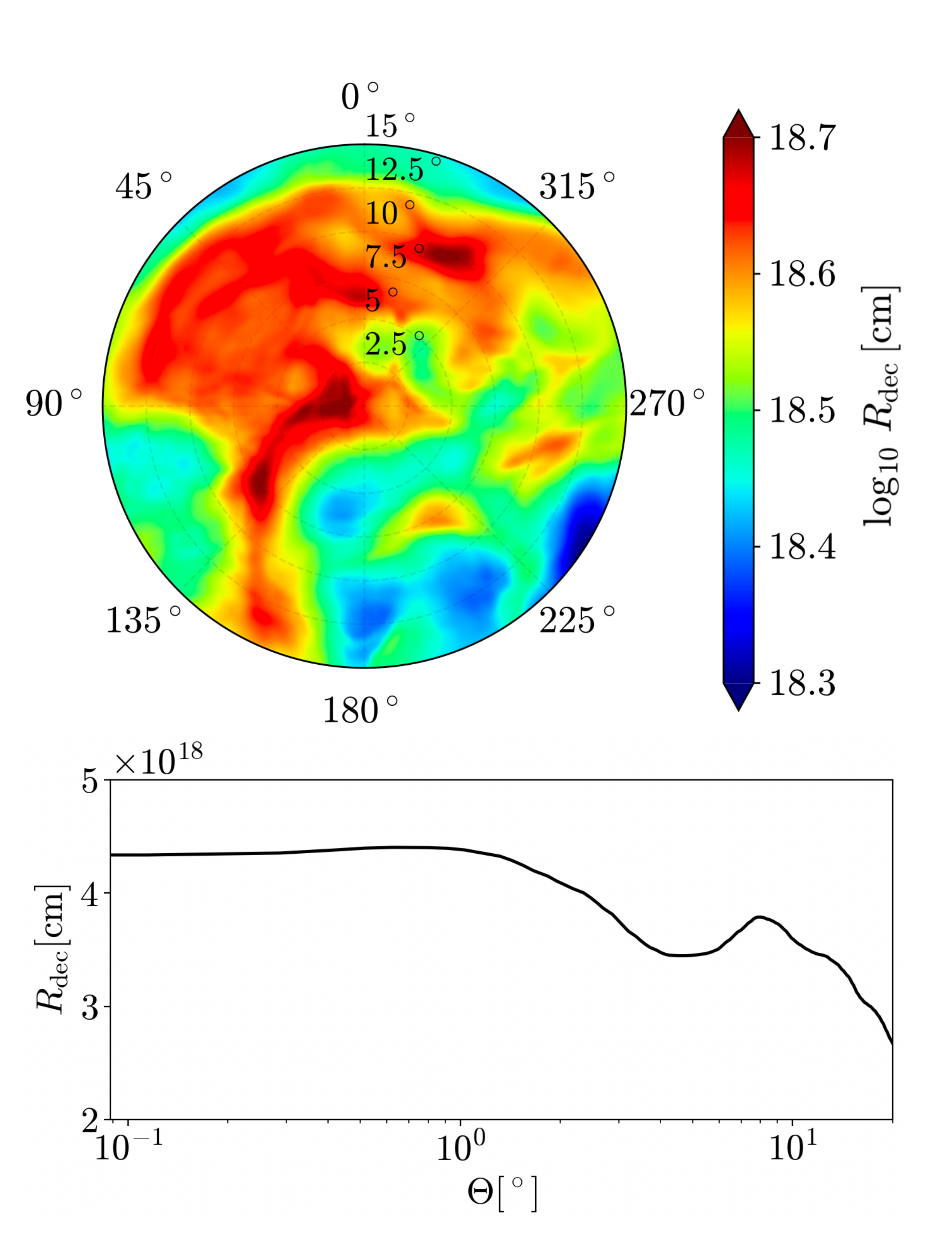}
    \caption{Angular profile of the deceleration radius estimated for our simulation, computed with Eq.~(\ref{eq:rd}).}
    \label{fig:dec}
\end{figure}

In Fig.~\ref{fig:dec} we show the results for $R_\mathrm{dec}$ obtained from the $E_\mathrm{iso}$ and $\bar{\Gamma}$ distributions at the end of the simulation, assuming $n=10^{-3}$\,cm$^{-3}$. The upper plot represents a front-view angular distribution, while the lower panel is obtained by using the azimuthally averaged profiles, showing the deceleration radius as a function of $\Theta$. Since the azimuthal symmetry assumption is most used in literature, we focus the following analysis on the lower panel.
We find $R_\mathrm{dec}(\Theta)$  $\sim [2.7-4.3]\times 10^{18}$~cm, barely consistent with the typical range $10^{16}-10^{18}$~cm for (long) GRBs \citep[]{M_sz_ros_2006,Meszaros1993b}. 
Typical values of $n$ for short GRBs have been constrained in the ample range $[10^{-5}-1]$\,cm$^{-3}$ (e.g. \citealt{Salafia2022}), thus allowing for values of $R_\mathrm{dec}$ from ten times smaller to a few times larger. 

From Fig.~\ref{fig:dec} we also note how the shape of the profile presents a ``valley'' at $\Theta$ around $4^\circ$.
This reflects the combination of a $\Phi$-averaged Lorentz factor that is peaked around that angle and a rather uniform  $\Phi$-averaged $E_\mathrm{iso}(\Theta)$ distribution in the range (1-4)$^\circ$ (see Fig. \ref{fig:angcfr}).

The corresponding deceleration time in the observer's frame can be estimated as $t_\mathrm{obs}(\Theta) = R_\mathrm{dec}(\Theta)/c \Gamma(\Theta)^2$, and ranges from $\sim 5 \times 10^5$\,s close to the jet injection axis to $\sim 5\times 10^6$\,s at large angles ($\Theta\approx 20^\circ$).

%%%%%%%%%%%%%%%%%%%%%%%%%%%%%%%%%%%%%%%%%%%%%%%%%%
\section{Summary and discussion}
\label{sec:conclusion}

We simulated the propagation of a short GRB jet emerging from a realistic BNS merger environment in 3D.
We presented the methods adopted to extend the simulation up to about 10 s, and analysed the dynamics, energetics, and structure of the outflow.
As a starting point, we used the final output of the simulation presented in \cite{Pavan2023}. 

In order to maintain a good resolution of the jet structure at large distances, at 3\,s from the jet launching time we switched from a spherical to a Cartesian computational domain, that allowed us to avoid any loss of resolution intrinsic to a spherical grid with logarithmic radial spacing.
The most relevant evolution of the jet properties and energetic conversions occur in the first three seconds of propagation, due to the significant interaction with the environment.
At later times, as result of the radial expansion, the high velocity part of the outflow evolves into a shell, with a final shape that significantly deviates from axisymmetry. 
In terms of energetics, we verified the conservation of the total energy in the Cartesian domain and followed the continuous conversion of magnetic and internal energies into kinetic, which reaches almost 98\% of the total by the end of the simulation.
We then investigated the radial and angular distributions of the most energetic portion of the jet, introducing a specific quantitative definition of the jet `head' and focussing on those properties that are most relevant to compute the electromagnetic emission, namely the isotropic equivalent energy and Lorentz factor. 
From this analysis, we confirmed that the jet angular distribution is essentially frozen towards the end of our evolution, which, together with the nearly complete energy conversion into kinetic form, makes our final outcome a reliable input for computing the afterglow emission.\footnote{We notice that this simulation does not consistently include radiative dissipation. Its possible role on the jet evolution goes beyond the aim of this work.}
Finally, we derived an estimate of the angular profile of the deceleration radius. Such a profile indicates that different portions of the outflow decelerate at different times, with possible visible effects on the afterglow light curves. 

We remark that, for the injection parameters employed here, the jet has to spend a large fraction of its energy to drill through the post-merger environment, with a final Lorentz factor limited to slightly more than 10. While such a jet is unlikely powerful enough to produce a canonical sGRB,  it still represents a possible physical case.
The systematic application to a number of different jet injection parameters (e.g. magnetisation, initial luminosity, among others), including cases corresponding to typical sGRB jets, will be the subject of future work. 
We further plan to compute afterglow emission light curves for each case and compare them with the observations of GRB\,170817A to plausibly obtain constraints on the physical conditions at the jet injection.

A remaining, overall limitation of the current framework is that the jet is manually injected into the realistic post-merger environment, instead of being consistently produced within the BNS merger simulation. 
Work is underway to overcome such a limit.

%%%%%%%%%%%%%%%%%%%%%%%%%%%%%%%%%%%%%%%%%%%%%%%%%%
\begin{acknowledgements}

We thank Om Sharan Salafia for insightful discussions. We further thank Giancarlo Ghirlanda and Andrea Mignone for their useful comments.
We acknowledge the referee for the constructive criticism that helped to significantly improve the paper.
This work was supported by the European Union under NextGenerationEU, via the PRIN 2022 Projects ``EMERGE: Neutron star mergers and the origin of short gamma-ray bursts", Prot. n. 2022KX2Z3B (CUP C53D23001150006), and ``PEACE: Powerful Emission and Acceleration in the most powerful Cosmic Explosion'', Prot. n. 202298J7KT (CUP G53D23000880006). The views and opinions expressed are solely those of the authors and do not necessarily reflect those of the European Union, nor can the European Union be held responsible for them.
AP and RC acknowledge further support by the Italian Ministry of Foreign Affairs and International Cooperation (MAECI), grant number US23GR08.  
Simulations were performed on the Discoverer HPC cluster at Sofia Tech Park (Bulgaria). We acknowledge EuroHPC Joint Undertaking for awarding us access to this cluster via the Regular Access allocations EHPC-REG-2022R03-218 and EHPC-REG-2023R03-160.

\end{acknowledgements}

%%%%%%%%%%%%%%%%%%%%%%%%%%%%%%%%%%%%%%%%%%%%%%%%%%
\section*{Data Availability}

The data underlying this article will be shared on reasonable request to the corresponding authors.


%\documentclass[bibyear]{aa}
%\bibpunct{(}{)}{;}{a}{}{,} % to follow the A&A style
%\bibliographystyle{aa} % style aa.bst
%\bibliography{refs} % your references Yourfile.bib

%%%%%%%%%%%%%%%%%%%%%%%%%%%%%%%%%%%%%%%%%%%%%%%%%%

%%%%%%%%%%%%%%%%% APPENDICES %%%%%%%%%%%%%%%%%%%%%

\begin{appendix}
\onecolumn

\section{Interpolation accuracy and jet head radial extension}

\subsection{Interpolation accuracy}
\label{sec:interpolation}

Here we quantify the impact of interpolating the simulation data from a spherical to a Cartesian grid (Sect. \ref{sec:setup}), 
and back to a new spherical grid (Sect. \ref{sec:head}), which is necessary to analyse the outcome of the evolution in terms of angular distributions.
In particular, we compare $\bar{\Gamma}(\Theta)$ and $E_{\mathrm{iso}}(\Theta)$ at $t-t_\mathrm{jet}=3$\,s obtained (i) from the original evolution in spherical coordinates and (ii) after applying a double interpolation spherical-to-Cartesian and Cartesian-to-spherical as employed in our work.
We note that the resolution of the Cartesian grid is slightly lower than the spherical one closer to the origin, and higher elsewhere (including within the head of the jet).

\begin{figure*}[h!]
    \centering   \subfigure{\includegraphics[width=0.335\textwidth]{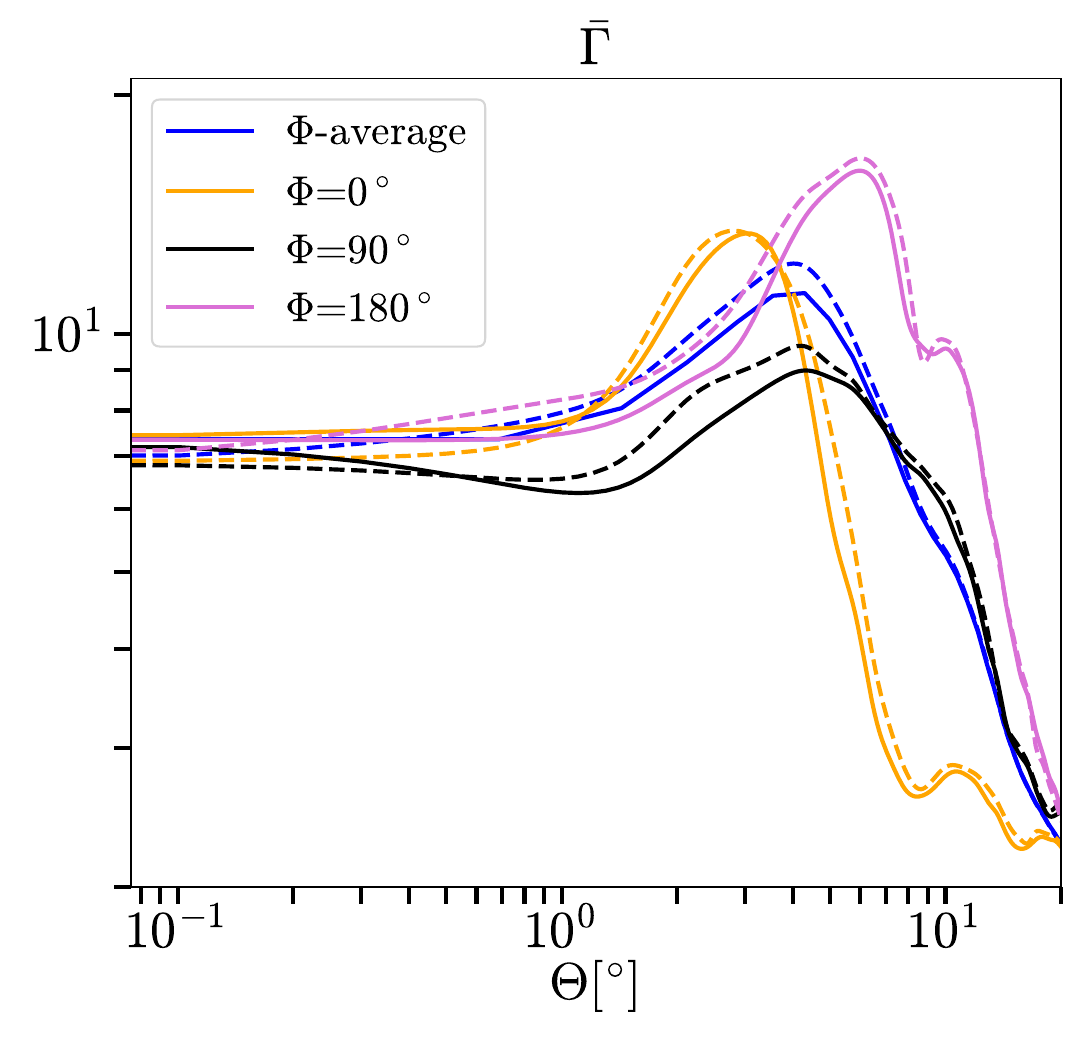}} 
    \subfigure{\includegraphics[width=0.34\textwidth]{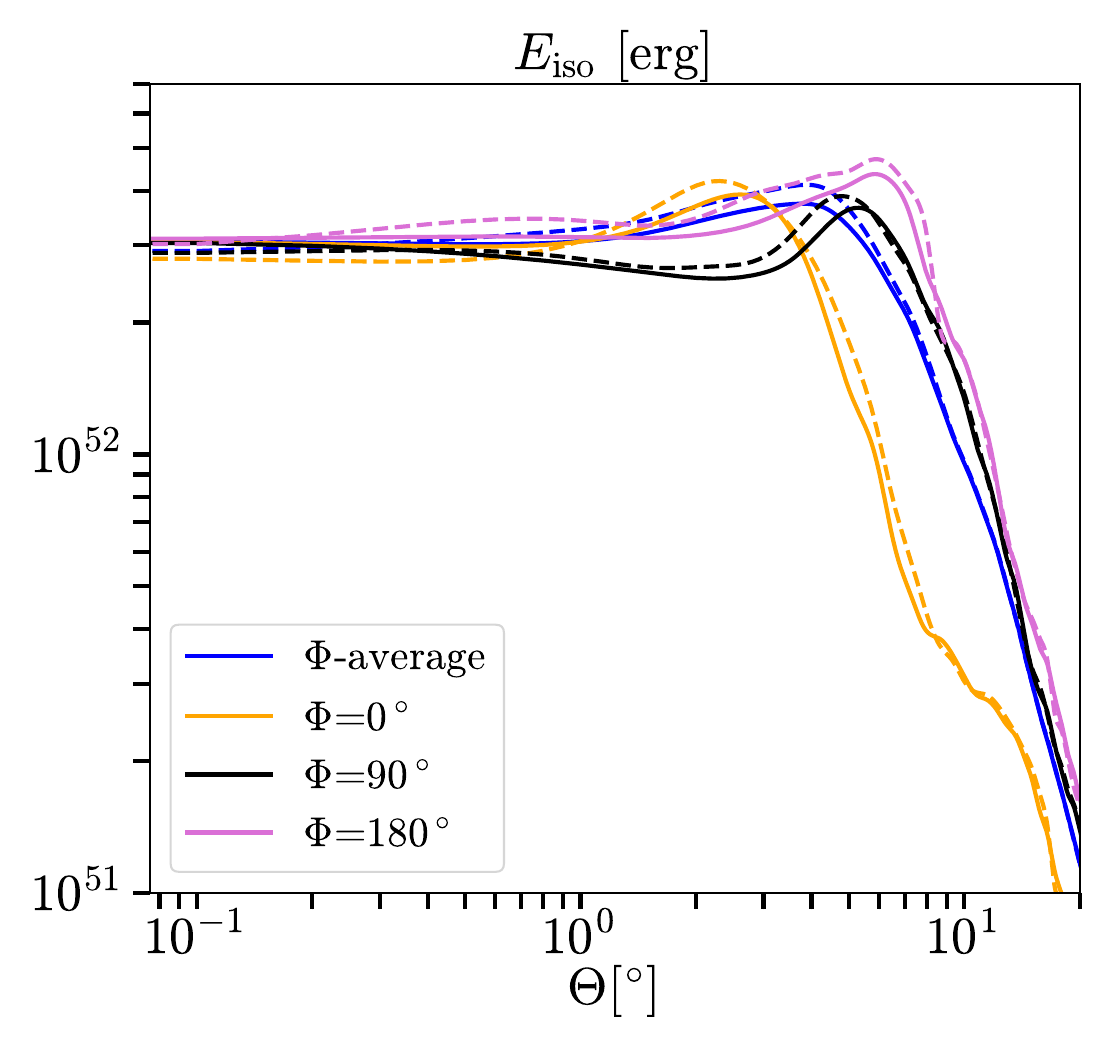}} 
    \begin{minipage}[b]{0.27\textwidth}
\caption{\small{Radially averaged Lorentz factor (top) and isotropic equivalent energy (bottom) of the jet head (see Sect.~\ref{sec:head}) as functions of $\Theta$, at $t-t_\mathrm{jet} \simeq $ 3\,s. We compare the result for the original input data (dashed lines) with those obtained after a double spherical-to-Cartesian and Cartesian-to-spherical interpolation (continuous lines). We refer to the text for further details.
    Different colours refer to profiles for three different azimuthal angles $\Phi=0^\circ,90^\circ,180^\circ$ (in yellow, black, purple lines, respectively) and for the $\Phi$-averaged result (blue line).}}
    \label{fig:domcfr}
\end{minipage}
    
\end{figure*}

The comparisons are shown in Fig. \ref{fig:domcfr}, for both $\bar{\Gamma}$ (top panel) and $E_{\mathrm{iso}}$ (bottom panel).
Four different profiles are considered (in different colours), referring to specific values of the azimuthal angle $\Phi$ and the $\Phi$-averaged result.
The resulting profiles are slightly different, as expected after a double interpolation changing coordinates and resolution, but still in rather good accordance. 
We find a maximum relative difference $\lesssim10$~\% for both quantities.
We remark that in the passage from the original spherical grid to the Cartesian one, which is used to continue the evolution after 3 s from the jet launching time, there is no loss of resolution affecting the results of the simulation.

To give a further estimate (less affected by the grid details) of the interpolation accuracy, we also compared the total energy contained within the same volume inside the computational domain.
The volume considered is a spherical section with $r\in [7 \times 10^5, 1.3 \times 10^6]$ \,km and a half-opening angle of 40$^\circ$ around the jet injection axis. 
As a result, we find a relative difference of about 2\%.

%%%%%%%%%%%%%%%%%%%%%%%%%%%%%%%%
\subsection{Jet head radial extension}
\label{sec:threshold}

\begin{figure*}[h!]
    \centering    \subfigure{\includegraphics[width=0.335\textwidth]{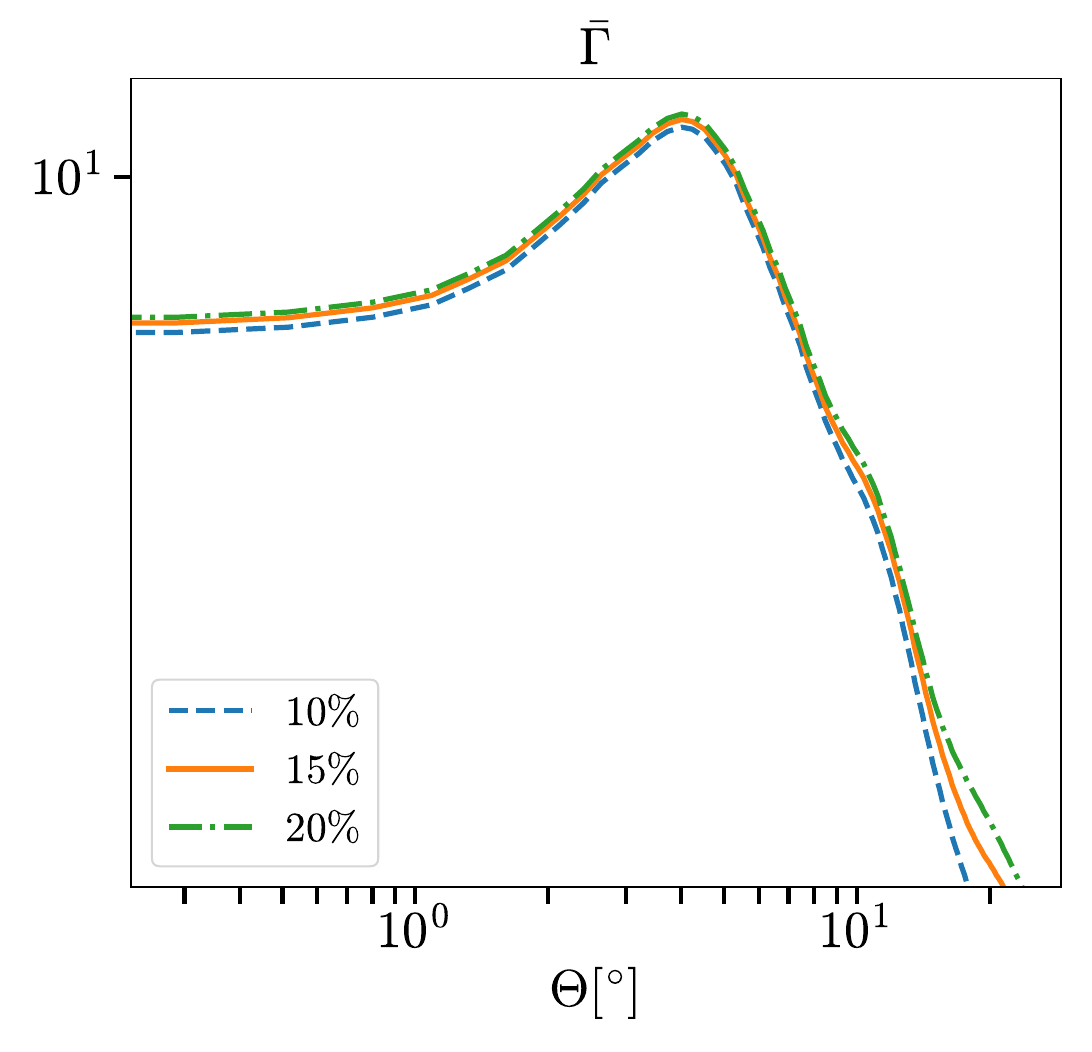}}     
    \subfigure{\includegraphics[width=0.34\textwidth]{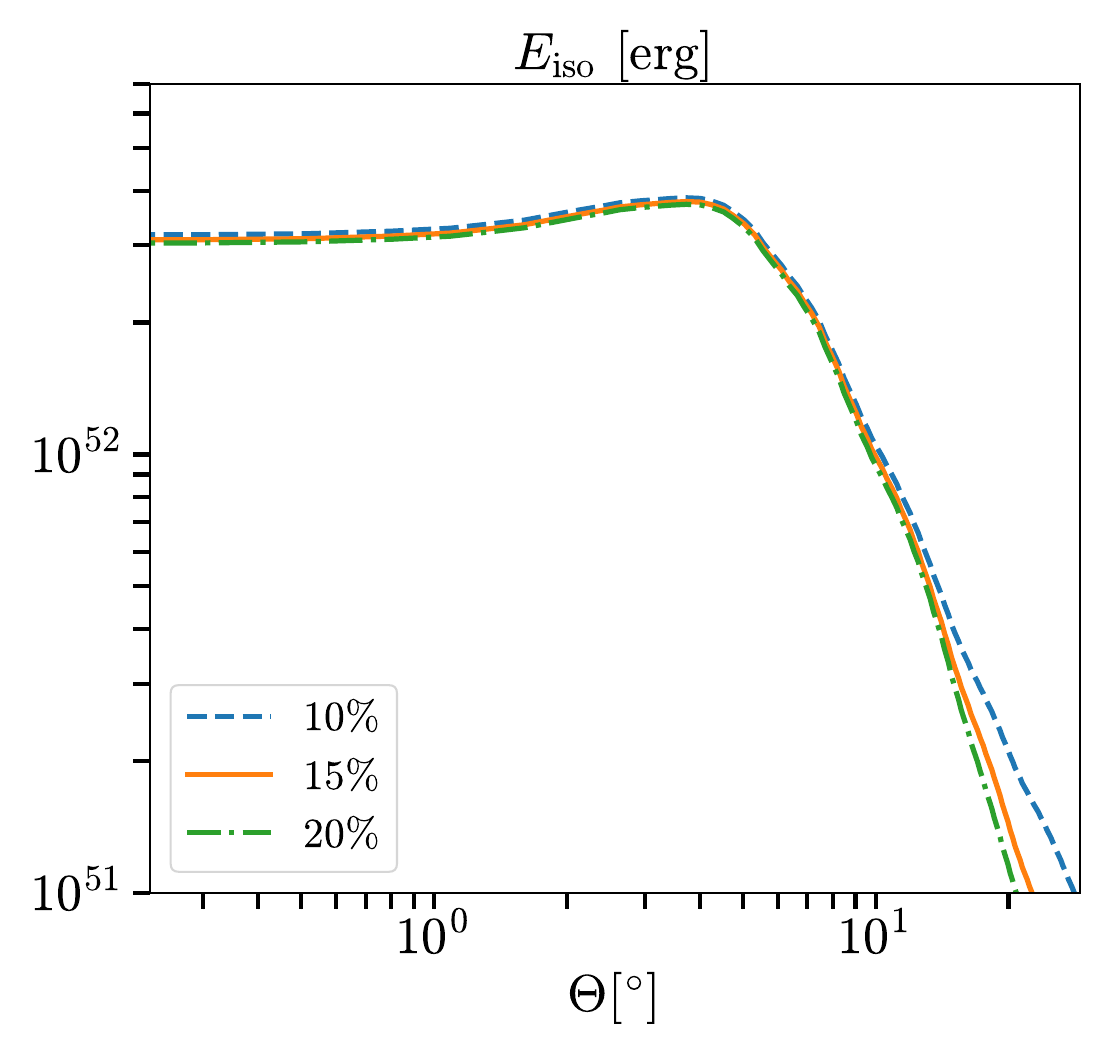}} 
    \begin{minipage}[b]{0.27\textwidth}   
\caption{\small{$\Phi$-averaged profiles of the radially averaged Lorentz factor (top) and isotropic equivalent energy (bottom) of the jet head as functions of $\Theta$ at time $t_1$, that is shortly after the beginning of our Cartesian simulation. The three profiles in each panel refer to the results obtained for different threshold fractions of the maximum $dE/dt$ adopted to define the radial extension of the jet head (see Sect. \ref{sec:head}), namely 10\%, 15\%, and 20\%. 
\\
\\}
    }
    \label{fig:treshcfr}
    \end{minipage}
\end{figure*}

We investigate the solidity of our definition for the jet head.
Figure \ref{fig:treshcfr} shows the $\Phi$-averaged profiles of $\bar{\Gamma}$ and $E_\mathrm{iso}$ at 3.5\,s after jet launching time, varying the threshold fraction of the maximum $dE/dt$ adopted to define the radial extension of the jet head (see Sect. \ref{sec:head}).
For both quantities, the profiles show that using a threshold of 10\%, 15\%, or 20\% gives very similar quantitative results, thus justifying our choice of 15\% as fiducial fraction.

\end{appendix}
%%%%%%%%%%%%%%%%%%%%%%%%%%%%%%%%%%%%%%%%%%%%%%%%%%

%\bsp	% typesetting comment
\label{lastpage}
\end{document}